\documentclass[onecolumn,preprint,superscriptaddress]{revtex4-1}
\usepackage{bibentry,natbib}
\usepackage{latexsym,bm}
\usepackage{graphicx}
\usepackage{float}
\usepackage{rotating}
\usepackage{amsmath}
\usepackage{rotating}
\usepackage{color}
\usepackage[colorlinks,citecolor=red]{hyperref}
\usepackage{subfigure}
\usepackage{ulem}
\DeclareGraphicsExtensions{.pdf}

\begin{document}
	\title{Investigation of Efimov Features and Universality in $^{87}$Rb-$^{40}$K Mixtures with finite-range interaction}
	\author{Ning-Ning Gao}
	\affiliation{State Key Laboratory of Magnetic Resonance and Atomic and Molecular Physics, Wuhan Institute of Physics and Mathematics, Innovation Academy for Precision Measurement Science and Technology, Chinese Academy of Sciences, Wuhan 430071, People's Republic of China}
	\affiliation{University of Chinese Academy of Sciences, Beijing 100049, People's Republic of China}
	\author{Hui-Li Han}
	\email{huilihan@wipm.ac.cn}
	\affiliation{State Key Laboratory of Magnetic Resonance and Atomic and Molecular Physics, Wuhan Institute of Physics and Mathematics, Innovation Academy for Precision Measurement Science and Technology, Chinese Academy of Sciences, Wuhan 430071, People's Republic of China}
	\author{Ting-Yun Shi}
\affiliation{State Key Laboratory of Magnetic Resonance and Atomic and Molecular Physics, Wuhan Institute of Physics and Mathematics, Innovation Academy for Precision Measurement Science and Technology, Chinese Academy of Sciences, Wuhan 430071, People's Republic of China}
	\date{\today}
	
	\begin{abstract}
    The study of Efimov features and their relationships in $^{40}$K-$^{87}$Rb Mixtures has generated extensive discussion, yet the discrepancy between Efimov universality predictions based on the zero-range approximation and experimental observations remains unresolved. In this study, we investigate the three-body collision properties with $J=0$ symmetry for a $^{87}$Rb-$^{87}$Rb-$^{40}$K system on both sides of Rb-K scattering length to understand the mechanisms underlying this discrepancy. Our approach employs the R-matrix propagation method within a hyperspherical coordinate frame, utilizing the Lennard-Jones model potential to describe atom interactions. We predicts the existence of three-body shape resonances at large negative Rb-K scattering lengths, which leads to the enhancement of three-body recombination rates. On the positive Rb-K scattering length side, we find an Efimov recombination minimum beyond the range of previous measurements. These Efimov features, combined with experimental observations of the atom-dimer Efimov resonance, offer an opportunity to test the universality of Efimov features. Our study demonstrates the influence of finite-range effects and non-resonant intraspecies scattering length in $^{40}$K-$^{87}$Rb mixtures, providing valuable insights into the universal relations between Efimov features in heteronuclear systems.
	\end{abstract}
	\maketitle
	\section{Introduction}
	The Efimov effect, originally predicted in 1970\;\cite{Efimov1970,Efimov1973}, has attracted broad interest in atomic and nuclear physics\,\cite{Kraemer2006,Zaccanti2009,Bloom2013,HUlmanis2016,
		Barontini2009,Jensen2004structure,Kolganova2011}. When the two-body s-wave scattering length $a$ is tuned to a large value compared to the characteristic range $r_{0}$ of the two-body interaction potential, an infinite series of trimer states known as Efimov trimers are formed\,\cite{Naidon2017}. These Efimov trimer states exhibit discrete scale invariance, characterized by the relation: $E_{n} = \lambda^{2}E_{n+1}$, where $n$ denotes the $n$th Efimov state with three pairs of interactions in the resonant limit\,\cite{Eric2006}. The scaling constant $\lambda $ is determined by $e^{\pi/s_{0}}$, where $s_0$ depends on the number of resonant interactions, as well as the quantum statistics and mass ratio of a trimer's constituents\,\cite{IncaoMass2006}.
	
\begin{figure}[htbp]
		\centering
		\subfigure{
			\includegraphics[width=9.0cm]{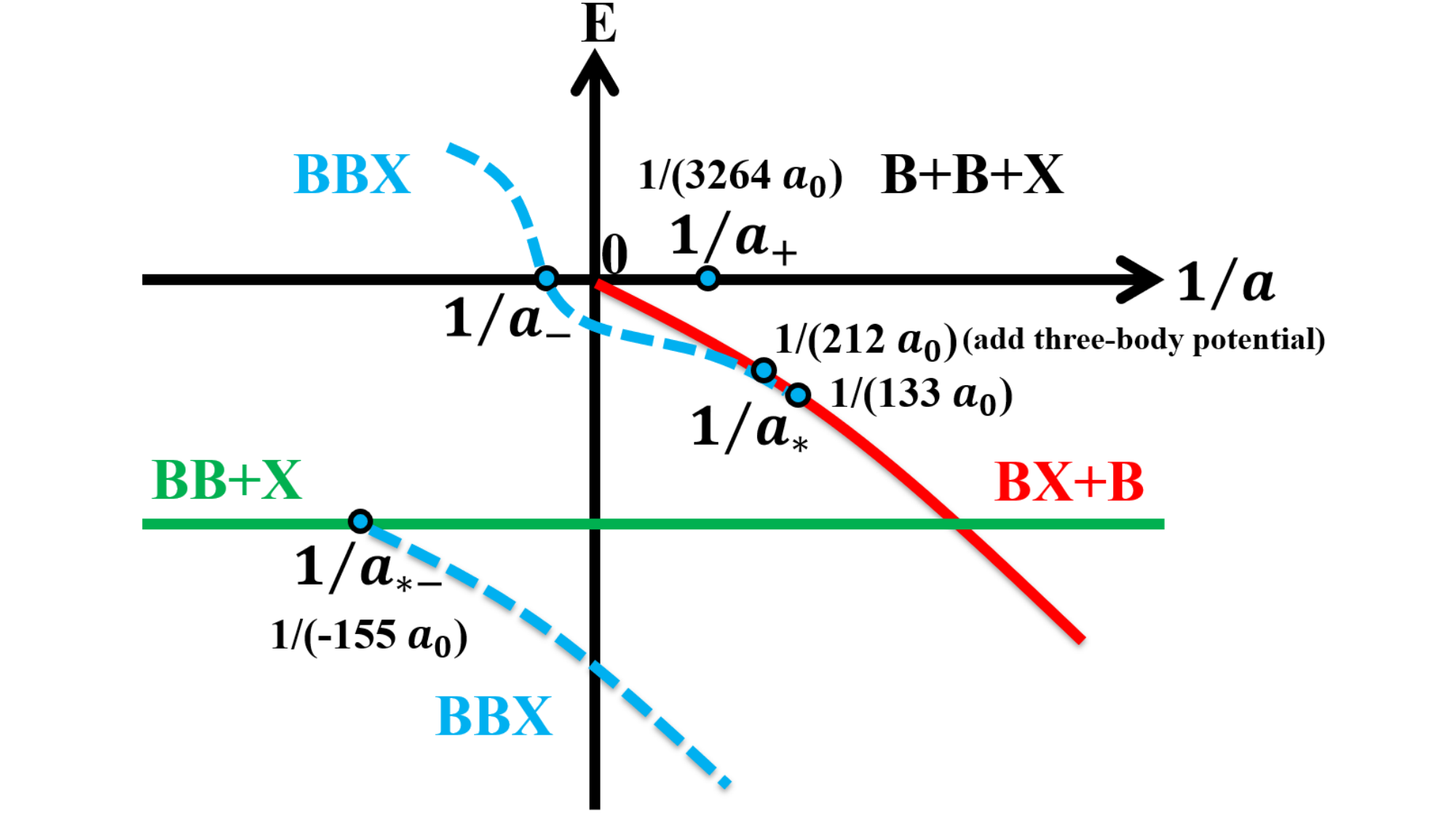}
		}
\caption{(Color online) Schematic representation of the Efimov features in the presence of a BB bound state. The thick black line denotes the threshold energy of three free atoms, the red solid line represents the threshold for a BX Feshbach molecule plus a free B atom, and the thick green line corresponds to the BB + X threshold. The blue dashed lines indicate the trimer states. The calculated positions of $a_{+}$, $a_{*}$, and $a_{*-}$ (the $^{40}$K-$^{87}$Rb$^{87}$Rb resonance positions) are also labeled.}
\label{fig1}
	\end{figure}

In ultracold gases, where the two-body scattering length $a$ can be tuned using Feshbach resonances\,\cite{Chin2010}, the Efimov effect is observed by measuring the three-body event rate constant\,\cite{D_Incao_2018}. This constant is resonantly enhanced when an Efimov trimer state approaches zero binding energy at a negative scattering length, denoted $a_{\scriptscriptstyle-}$. On the other side of the Feshbach resonance, interference effects cause minimum in the three-body rate coefficient at a positive scattering length, $a_{\scriptscriptstyle+}$, where an Efimov trimer state coincides with the atom-molecule threshold, denoted $a_{\scriptscriptstyle*}$. These Efimov features are illustrated in Fig.\,\ref{fig1} and are commonly referred to as three-body parameters (3BP).

From an experimental perspective, one direct way to identify the Efimov effect in ultracold quantum gases is by observing multiple consecutive Efimov features in atomic and molecular losses\,\cite{Maier2015,Tung2014Dec,HUlmanis2016,Huang2014,hafner2017role}, which follow a characteristic geometric scaling\,\cite{D_Incao_2018}. However, this approach requires tuning the scattering length to very large values and achieving extremely low temperatures to maintain the system within the threshold regime\,\cite{d2004limits}. Currently, only a few systems exhibit two or more Efimov features\,\cite{Tung2014Dec,HUlmanis2016}, which are necessary to test universal relations.

There are, however, other Efimov-related properties that may bypass the need for such extreme conditions\,\cite{d2004limits}. The observation of a single Efimov feature in any two observables and the verification of their corresponding universal relations provide an alternative way to demonstrate the physics of the Efimov effect\,\cite{WangJia2012,NaidonAug2014,NaidonMar2014}. The universal relations between Efimov features in different scattering observables were first determined by Braaten and Hammer in Ref.\,\cite{Eric2006}, with later studies incorporating range corrections to refine these universal relationships\,\cite{Gogolin2008Apr}. Mestrom \textit{et al.}\,\cite{mestrom2017efimov} tested these universal relationships and explored deviations using a finite-range interaction model with positive scattering length in a homonuclear system.

However, these relationships become more complex in mixed BBX systems, where two heavy atoms B resonate with a lighter atom X due to the presence of two different scattering lengths\,\cite{Wacker2016KRb,WangYujun2012}. Near the interspecies Feshbach resonance, interactions between the two identical atoms, governed by the smaller scattering length, can lead to deviations from standard scaling laws\,\cite{D_Incao_2018,Helfrich2010}. Furthermore, for positive intraspecies scattering lengths, the X+BB channel can split the potential curves into two Efimov branches, complicating the universal relationship even further\,\cite{HUlmanis2016,hafner2017role}. As a result, investigating universal relationships between Efimov features in such systems remains a compelling and important topic of study.

The Efimov effect in the K-Rb mixed system has garnered considerable attention in both theoretical and experimental research\,\cite{WangYujun2012,Bloom2013}. Several experimental groups have investigated Efimov-unfavored systems involving K-Rb mixtures\,\cite{Barontini2009,Hu2014,Wacker2016KRb}. The first experiment with $^{41}$K-$^{87}$Rb Bose-Bose mixtures revealed an unexpected non-universal behavior of Efimov resonances, with $a_{\scriptscriptstyle-} = -246\,a_{0}$, where $a_{0}$ is the Bohr radius\,\cite{Barontini2009}. This finding spurred additional experiments on $^{40}$K-$^{87}$Rb Fermi-Bose mixtures, which consistently showed an Efimov-type resonance near $a_{\scriptscriptstyle*} = 230(30)\,a_{0}$ in atom-dimer collisions\,\cite{Zirbel2008,Bloom2013}.

The predicted universal relationship for $^{40}$K-$^{87}$Rb mixtures, assuming $a_{\scriptscriptstyle\textsl{RbRb}} = 0$ based on effective-field theory, is $|a_{\scriptscriptstyle-}|/a_{\scriptscriptstyle*} = 240$\,\cite{D_Incao_2018}. Given the observed resonance at $a_{\scriptscriptstyle*} = 230(30)\,a_{0}$, this relationship would suggest $a_{\scriptscriptstyle-} = -55200\,a_{0}$, which lies beyond the reach of current experiments. For positive scattering lengths in the $^{40}$K-$^{87}$Rb system, the predicted recombination minimum at $a_{\scriptscriptstyle+} = 2800\,a_{0}$, as proposed by Ref.\,\cite{WangYujun2012}, was ruled out by subsequent experiments\,\cite{Bloom2013}. Furthermore, the estimate of Efimov resonances with $a_{\scriptscriptstyle-} < -30000\,a_{0}$ given by Ref.\,\cite{WangYujun2012} lacks the precision necessary to thoroughly test the universal relationship.

Based on the above analysis, this paper aims to calculate the Efimov features in $^{40}$K-$^{87}$Rb Fermi-Bose mixtures by investigating the three-body recombination and atom-dimer collision processes, with the $^{87}$Rb-$^{87}$Rb interaction fixed at $a_{\scriptscriptstyle\textsl{RbRb}} = 100\,a_{0}$, which is close to the experimental condition. The goal of this study is to clarify the existence of two Efimov branches in the K-Rb system and provide accurate Efimov feature $a_{\scriptscriptstyle+}$ for the system. We use the R-matrix propagation method in the hyperspherical coordinate frame based on the Lennard-Jones model potential for the interactions between atoms as described in Ref.\,\cite{WangJia2011}. The hyperradius is divided into two regions: short distances, where the slow-variable-discretization (SVD) method\,\cite{Tolstikhin1996SVD} is applied to overcome the numerical difficulties at sharp nonadiabatic avoided crossings; and large distances, where the traditional adiabatic hyperspherical method is employed to avoid the large memory and central processing unit time needed in SVD. Then, the $\underline{\mathcal{R}}$ matrix was propagated from short to large distances. The scattering properties are obtained through the $\underline{\mathcal{S}}$ matrix by matching the $\underline{\mathcal{R}}$ matrix with asymptotic functions and boundary conditions\,\cite{WangJia2011}.
	
	The paper is organized as follows.
	In Sec. II, our calculation method and all necessary formulas for calculations are presented.
	In Sec. III, we discuss the results and the universal relationship between the different Efimov features.
	Finally, we provide a brief summary. Atomic units are
	applied throughout the paper unless stated otherwise. \\
	
	\section{Theoretical formalism}

In this study, we investigate the collision properties of the $^{87}$Rb-$^{87}$Rb-$^{40}$K system in $J^{\Pi}=0^{+}$ symmetry. The masses of the three atoms, $^{87}$Rb, $^{87}$Rb, and $^{40}$K, are denoted by $m_{1}$, $m_{2}$, and $m_{3}$, respectively. We employ Delves's hyperspherical coordinates and introduce the mass-scaled Jacobi coordinates. The first Jacobi vector $\vec{\rho}_{1}$ is chosen to be the vector from atom 1 to atom 2, with reduced mass $\mu_{1}$, and the second Jacobi vector $\vec{\rho}_{2}$ goes from the diatom center of mass to the third atom, with reduced mass $\mu_{2}$. The angle between $\vec{\rho}_{1}$ and $\vec{\rho}_{2}$ is denoted by $\theta$. The hyperradius $\rho$ and hyperangle $\phi$ are defined as\\
	\begin{equation}
		\label{1}
		\mu \rho^{2}=\mu_{1}\rho_{1}^{2}+\mu_{2}\rho_{2}^{2}\,,
	\end{equation}
	and\\
	\begin{equation}
		\label{2}
		\tan\phi=\sqrt{\frac{\mu_{2}}{\mu_{1}}}\frac{\rho_{2}}{\rho_{1}},\;\; 0 \leq\phi\leq\frac{\pi}{2}\,,
	\end{equation}
	respectively, where $\rho$ is the only coordinate with the dimension of length and represents the overall size of the three-body system. The rotation of the plane containing the three particles is described collectively by $\Omega$ $[\Omega \equiv (\theta, \phi, \alpha, \beta, \gamma)]$, which includes $\theta$, $\phi$, and three Euler angles $(\alpha, \beta, \gamma)$. The parameter $\mu$ is an arbitrary scaling factor, and we choose $\mu=\sqrt{\mu_{1}\mu_{2}}$ for our calculations. \\
	
The Schr$\mathrm{\ddot{o}}$dinger equation in hyperspherical coordinates can be written after rescaling the three-body wave function $\Psi_{\upsilon'}$ as $\psi_{\upsilon'}(\rho;\theta,\phi)=\Psi_{\upsilon'}(\rho;\theta,\phi) \rho^{5/2} \sin\phi \cos\phi$:\\
	\begin{equation}
		\label{3}
		\left[ -\frac{1}{2\mu}\frac{d^{2}}{d\rho^{2}}+\left( \frac{\Lambda^{2}-\frac{1}{4}}{2\mu \rho^{2}}+V(\rho;\theta,\phi)\right) \right] \psi_{\upsilon'}(\rho;\Omega)=E\psi_{\upsilon'}(\rho;\Omega)\,,
	\end{equation}
	where $\Lambda^{2}$ is the squared ``grand angular momentum operator", and its expression is as given in Ref.\,\cite{lin1995}. The three-body interaction potential $V(\rho;\theta,\phi)$ is expressed as a sum of three pairwise two-body interaction potentials $v(r_{ij})$:\\
	\begin{equation}
		\label{4}
		V(\rho;\theta,\phi)=v(r_{12})+v(r_{13})+v(r_{23})\,,
	\end{equation}
	where $r_{ij}$ is the interparticle distance. Here, we use the Lennard-Jones potential to model the interactions between two atoms\,\cite{WangJia2012}, which was proven to be an excellent model potential to explore van der Waals universality in Efimov physics\;\cite{WangYujun2012,WangJia2012,HUlmanis2016,NaidonAug2014}. The potential is expressed in the form of \\
	\begin{equation}
		\label{5}
		v(r_{ij})=-\frac{C_{6,ij}}{r_{ij}^{6}}\left[1-\frac{1}{2}\left(\frac{\lambda_{ij}}{r_{ij}}\right)^{6}\right]\,,
	\end{equation}
	where the $i$ and $j$ indices in $r_{ij}$ label particles $i$ and $j$. The two-body scattering length $a_{ij}$ is changed by adjusting the $\lambda_{ij}$, and $C_{6,ij}$ is the dispersion coefficient. The values of $C_{6,RbRb}$ and $C_{6,RbK}$ used here are 4698 from Ref.\;\cite{C6RbRb2014} and 4106.5 from Ref.\,\cite{C6RbK1999}, respectively.\\
	
	The three-body wave function $\psi_{\upsilon^{'}}$ can be expanded with the complete, orthonormal set of angular wave function $\Phi_{\nu}$ and the radial discrete variable representation (DVR) basis $\pi_i(\rho)$ as\\
	\begin{align}
\label{6}
 \psi_{\upsilon'}(\rho;\Omega)=\sum^{N_{DVR}}_i\sum^{N_{chan}}_{\nu}C^{ \upsilon'}_{i\nu}\pi_i(\rho)\Phi_\nu(\rho_i;\Omega)\,.
\end{align}

	The adiabatic potentials $U_{\nu}(\rho)$ and channel functions $\Phi_{\nu}(\rho;\varOmega)$ at fixed $\rho$ can be obtained by solving the following adiabatic eigenvalue equation:\\
	\begin{equation}
		\label{7}
		\left( \frac{\Lambda^{2}-\frac{1}{4}}{2\mu \rho^{2}}+V(\rho;\theta,\phi)\right) \Phi_{\nu}(\rho;\varOmega)=U_{\nu}(\rho) \Phi_{\nu}(\rho;\varOmega)\,.
	\end{equation}

The hyperradius $\rho$ is divided into $(N-1)$ intervals using a set of grid points $\rho_1<\rho_2<\cdots \rho_N$. At short distances in the interval $[\rho_i,\rho_{i+1}]$, we utilize the SVD method to solve Eq.\,(\ref{3}) as a standard algebraic problem for the coefficients $C^{\upsilon'}_{i\nu}$:
\begin{align}
\label{8}
 \sum^{N_{DVR}}_{j}\sum^{N_{chan}}_{\mu}\mathcal{T}_{ij}
 \mathcal{O}_{i\nu,j\mu}C^{\upsilon'}_{j\mu}+U_\nu (\rho_i)C^{\upsilon'}_{i\nu}=E^{\upsilon'}C^{\upsilon'}_{i\nu}\,,
\end{align}
where
\begin{align}
\label{9}
\mathcal{T}_{ij}=\frac{1}{2\mu}\int^{\rho_{i+1}}_{\rho_{i}}\frac{d}{d\rho}\pi_i(\rho)\frac{d}{d\rho}\pi_j(\rho)d\rho\,,
\end{align}
are the kinetic energy matrix elements, with ${\rho_{i}}$ and ${\rho_{i+1}}$ being the boundaries of the calculation box, and
\begin{align}
\label{10}
\mathcal{O}_{i\nu,j\mu}=\langle\Phi_\nu(\rho_i;\Omega)|\Phi_\mu(\rho_j;\Omega)\rangle\,.
\end{align}
are the overlap matrix elements between the adiabatic channels defined at different quadrature points.

At large distances, the traditional adiabatic hyperspherical method is used to solve Eq.\,(\ref{3}).
When substituting the wave functions $\psi_{\upsilon'}(\rho;\Omega)$ into Eq.\,(\ref{3}), one obtains a set of coupled ordinary differential equations:
\begin{align}
\label{11}
[-\frac{1}{2\mu}\frac{d^2}{d\rho^2}+U_\nu(\rho)- E]F_{\nu,\upsilon'}(\rho)
-\frac{1}{2\mu}\sum_{\mu}[2P_{\mu\nu}(\rho)\frac{d}{d\rho}+Q_{\mu\nu}(\rho)]F_{\mu \upsilon'}(\rho)=0\,,
\end{align}
where
\begin{align}
\label{12}
P_{\mu\nu}(\rho)=\int d\Omega \Phi_{\mu}(\rho;\Omega)^{*}\frac{\partial}{\partial \rho}\Phi_{\nu}(\rho;\Omega)\,,
\end{align}
and
\begin{align}
\label{13}
Q_{\mu\nu}(\rho) = \int d\Omega \Phi_{\mu}(\rho;\Omega)^{*}\frac{\partial^{2}}{\partial \rho^{2}}\Phi_{\nu}(\rho;\Omega)\,.
\end{align}
are the nonadiabatic couplings that control the inelastic transitions as well as the width of the resonance supported by adiabatic potential $U_\nu(\rho)$.

The effective hyperradial potentials that include hyperradial kinetic energy contributions with the $P_{\nu\nu}^2$ term, are more physical than adiabatic hyperpotentials and are defined as
\begin{align}
\label{14}
W_{\nu \nu}(\rho)=U_{\nu}(\rho)-\frac{\hbar^{2}}{2 \mu} P_{\nu \nu}^{2}(\rho)\,.
\end{align}
The effective potentials for the recombination channels (labeled by $f$) have asymptotic behavior at large $\rho$ as\\
	\begin{equation}
		\label{15}
		W_{f}(\rho)=\frac{l_{f}(l_{f}+1)}{2\mu \rho^{2}}+E_{2b}\,,\\
	\end{equation}
	where $E_{2b}$ is the two-body bound state energy and $l_{f}$ is the corresponding angular momentum of the third atom relative to the dimer. For the three-body continuum channels (labeled by $i$), in which the three atoms are far away from each other as $\rho \to \infty$. The corresponding potential curves behave as\\
	\begin{equation}
		\label{16}
		W_{i}(\rho)=\frac{\lambda_i(\lambda_i+4)+15/4}{2\mu \rho^{2}}\,.\\
	\end{equation}
	Here, $\lambda_i(\lambda_i+4)$ is the eigenvalue of the grand angular momentum operator $\Lambda^{2}$, and $\lambda_i$ are nonnegative integer values, which are restricted by the requirements of permutation symmetry.\\

After we obtain the three-body wave function $\psi_{\upsilon'}(\rho;\Omega)$ and channel functions $\Phi_{\nu}(\rho;\varOmega)$ , the $\underline{\mathcal{R}}$ matrix
,
\begin{align}
\label{17}
\underline{\mathcal{R}}(\rho)=\underline{\textsl{F}}(\rho)[\widetilde{\underline{\textsl{F}}}(\rho)]^{-1}\,,
\end{align}
 can be calculated with $\psi_{\upsilon'}(\rho;\Omega)$ and $\Phi_{\nu}(\rho;\varOmega)$ by evaluating the integrals given below:
\begin{align}
\label{18}
F_{\nu,\upsilon'}(\rho)=\int d\Omega \Phi_{\nu}(\rho;\Omega)^{*}\psi_{\upsilon'}(\rho;\Omega)\,,
\end{align}
\begin{align}
\label{19}
\widetilde{F}_{\nu,\upsilon'}(\rho)=\int d\Omega \Phi_{\nu}(\rho;\Omega)^{*}\frac{\partial}{\partial \rho}\psi_{\upsilon'}(\rho;\Omega)\,.
\end{align}

Over an interval $\rho \epsilon [a_{1},a_{2}]$, for a given $R$-matrix at $\rho=a_{1}$, one uses the $R$-matrix propagation method to calculate the corresponding $R$-matrix at another point $\rho=a_{2}$ as follows,
\begin{align}
\label{20}
\underline{\mathcal{R}}(a_{2})=\underline{\mathcal{R}}_{22}
-\underline{\mathcal{R}}_{21}\left[\underline{\mathcal{R}}_{11}
+\underline{\mathcal{R}}(a_{1})\right]^{-1}\underline{\mathcal{R}}_{12}\,.
\end{align}
The definition of matrix elements is as follows:
\begin{align}
\label{21}
(\underline{\mathcal{R}}_{11})_{\nu\mu} = \sum\limits_{n }\frac{u_{\nu}^{(n)}(a_{1})u_{\mu}^{(n)}(a_{1})}{2\mu (\varepsilon_{n}-E)}\,,
\end{align}
\begin{align}
\label{22}
(\underline{\mathcal{R}}_{12})_{\nu\mu} = \sum\limits_{n }\frac{u_{\nu}^{(n)}(a_{1})u_{\mu}^{(n)}(a_{2})}{2\mu (\varepsilon_{n}-E)}\,,
\end{align}
\begin{align}
\label{23}
(\underline{\mathcal{R}}_{21})_{\nu\mu} = \sum\limits_{n }\frac{u_{\nu}^{(n)}(a_{2})u_{\mu}^{(n)}(a_{1})}{2\mu (\varepsilon_{n}-E)}\,,
\end{align}
\begin{align}
\label{24}
(\underline{\mathcal{R}}_{22})_{\nu\mu} = \sum\limits_{n }\frac{u_{\nu}^{(n)}(a_{2})u_{\mu}^{(n)}(a_{2})}{2\mu (\varepsilon_{n}-E)}\,,
\end{align}
here $\nu$ and $\mu$ denote different channels respectively and more details can referred in Ref.\,\cite{WangJia2011}.
Once we have the $\underline{\mathcal{R}}$ at the large distance, the $\underline{\mathcal{K}}$ matrix and $\underline{\mathcal{S}}$ can be expressed in the following matrix equation:
\begin{align}
\label{25}
\underline{\mathcal{K}}=
(\underline{f}-\underline{f}'\underline{\mathcal{R}})
(\underline{g}-\underline{g}'\underline{\mathcal{R}})^{-1}\,,
\end{align}

\begin{align}
\label{26}
\underline{\mathcal{S}}=(\underline{1} + i \underline{\mathcal{K}} )(\underline{1} - i \underline{\mathcal{K}} )\,,
\end{align}
where $f_{\nu \nu'}=\sqrt{\frac{2\mu k_\nu}{\pi}} \rho j_{l_\nu}(k_\nu \rho)\delta_{\nu \nu'}$ and $g_{\nu \nu'}=\sqrt{\frac{2\mu k_\nu}{\pi}} \rho n_{l_\nu}(k_\nu \rho)\delta_{\nu \nu'}$ are the diagonal matrices of energy-normalized spherical Bessel and Neumann functions.
For the recombination channel, $l_\nu$ is the angular momentum of the third atom relative to the dimer, and $k_\nu$ is given by $k_{\nu}=\sqrt{2 \mu\left(E-E_{2 b}\right)}$.
For the entrance channel, $l_{\nu}=\lambda_{\nu}+3 / 2$, and $k_{\nu}=\sqrt{2 \mu E}$.
The cross-section for atom-dimer scattering is expressed in terms of the $S$ matrix as\\
	\begin{equation}
		\label{27}
		\sigma_{2}=\sum\limits_{J,\Pi}\sigma_{2}^{J,\Pi}=\sum\limits_{J,\Pi}\frac{(2J+1)\pi}{k_{ad}^{2}}|S_{f\leftarrow i}^{J,\Pi}-1|^{2}\,,\\
	\end{equation}
	where $\sigma_{2}^{J,\Pi}$ is the partial atom-dimer elastic scattering cross section corresponding to the $J^{\Pi}$ symmetry, $k_{ad}=\sqrt{2\mu_{ad}(E-E_{2b})}$ is the atom-dimer wavenumber, $\mu_{ad}$ is the atom-dimer reduced mass, and $E_{2b}$ is the binding energy of the weakly bound molecular state.
	The atom-dimer scattering length is defined as\\
	\begin{equation}
		\label{28}
		a_{ad}=-\lim\limits_{k_{ad} \to 0}\frac{\tan\delta_{0}}{k_{ad}}\,,\\
	\end{equation}
		where $\delta_{0}$ is the phase shift for atom-dimer elastic scattering and is related to the diagonal $S$-matrix element by the formula\\
	\begin{equation}
		\label{29}
		S_{f\leftarrow i}^{0+}=\exp{(2i\delta_{0})}\,.\\
	\end{equation}
	\section{Results and Discussion}
    \subsection{On the negative side of the $^{87}$Rb-$^{40}$K scattering length}
	\subsubsection{Atom($^{40}$K) - dimer($^{40}$Rb$^{40}$Rb) elastic scattering }

\begin{figure}[htbp]
		\centering
		\subfigure{
			\includegraphics[width=5.0cm]{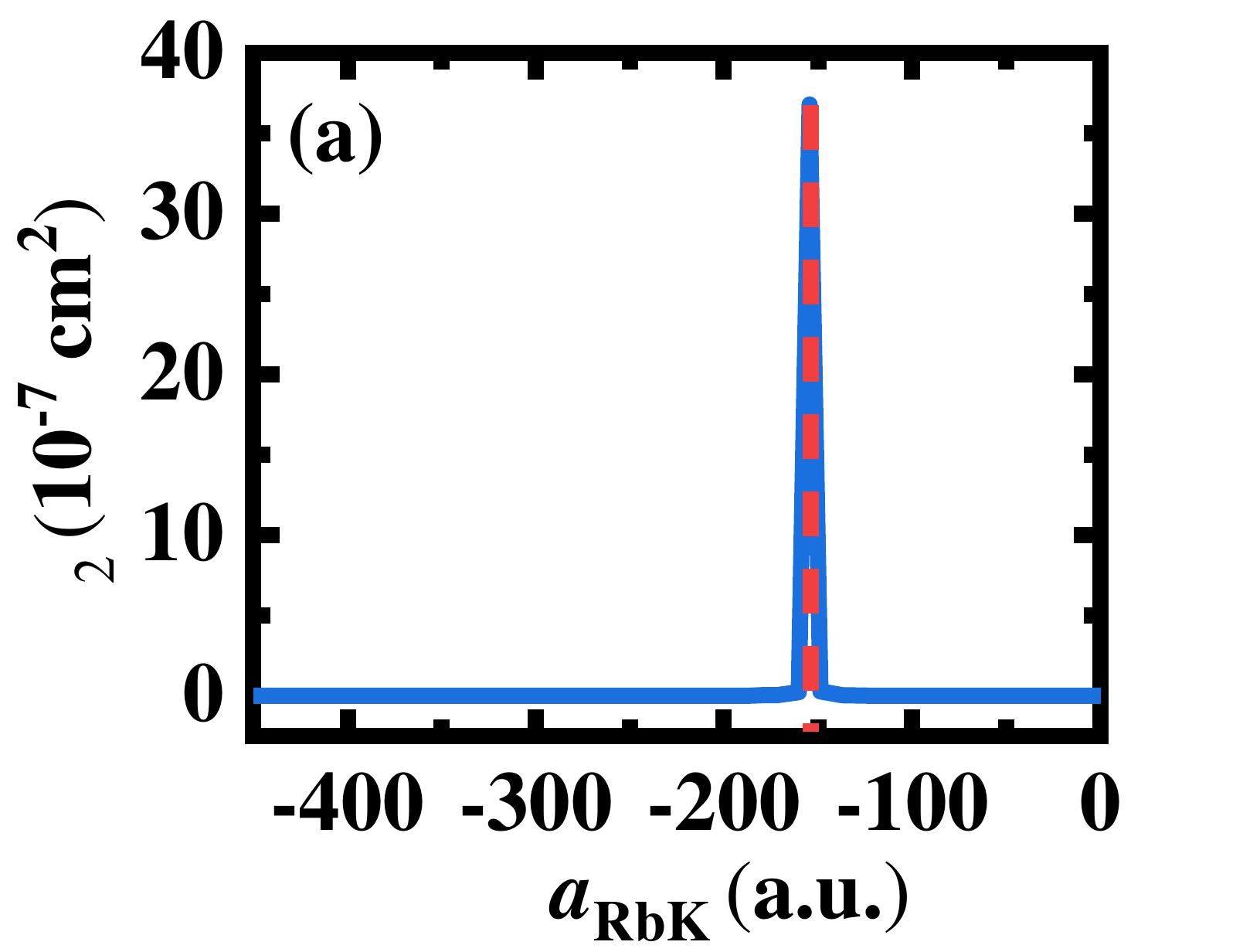}
			\label{fig2a}
		}
		\subfigure{
			\includegraphics[width=5.0cm]{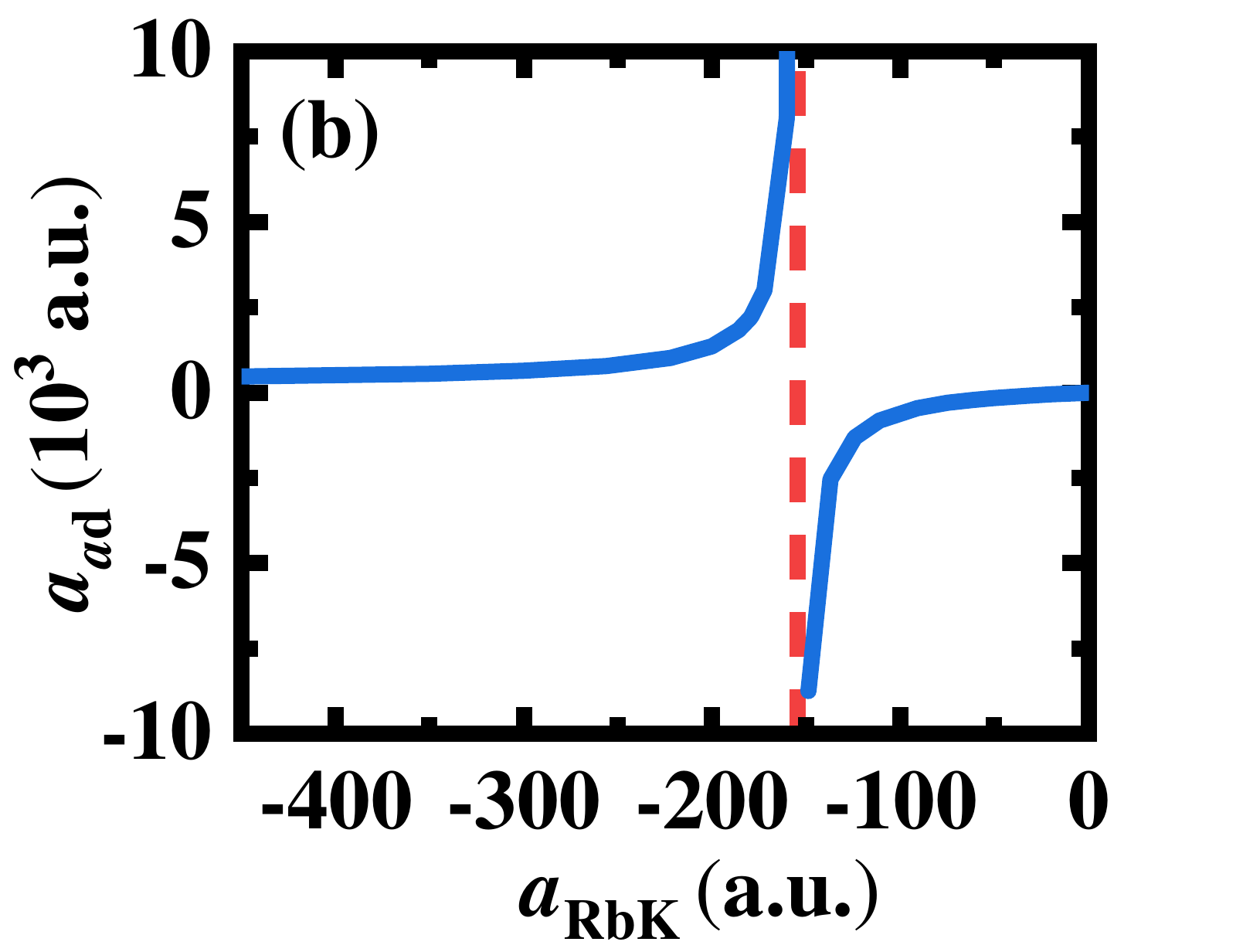}
			\label{fig2b}
		}
		\subfigure{
			\includegraphics[width=5.4cm]{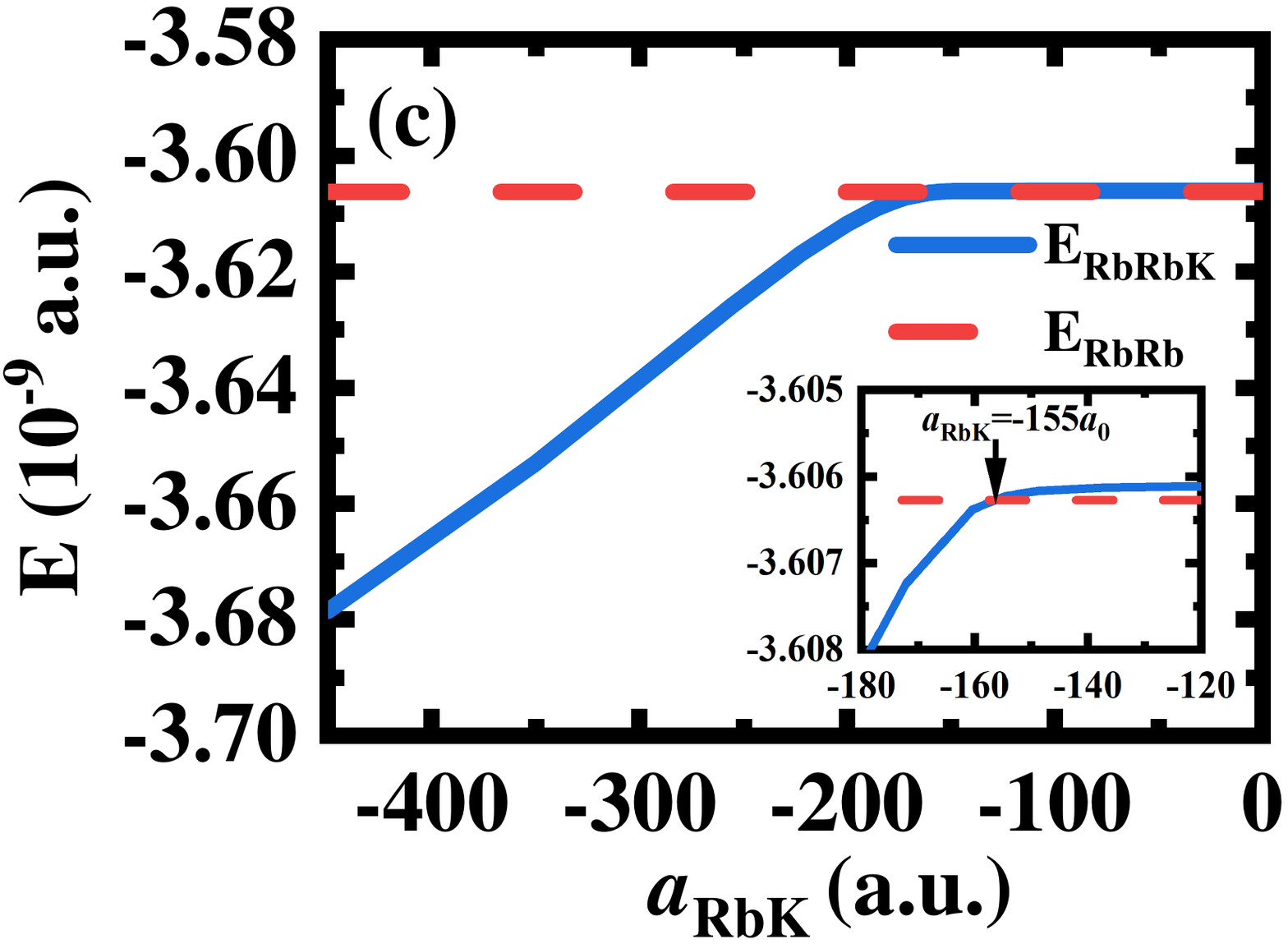}
			\label{fig2c}
		}
		\caption{(Color online)
			The $^{40}$K + $^{87}$Rb$^{87}$Rb elastic scattering observables with the $J^{\Pi}=0^{+}$ symmetry as a function of Rb-K scattering lengths when $a_{\scriptscriptstyle\textsl{RbRb}} = 100\,a_{0}$.
			All the data are obtained at low collision energy $E=100$ nK. (a) Cross sections: the red dashed line shows the position of the peak; (b) the atom-dimer scattering lengths; and the red dashed line shows the divergence position; (c) three-body energy spectrum. The red dashed line is the RbRb energy, and the solid line is the three-body state energy. The inset shows that the three-body state disappears at $a_{\scriptscriptstyle\textsl{RbK}}=-155\,a_{0}$.}
		\label{fig2}
	\end{figure}

	In this section, our focus is on the scenario where $^{87}$Rb-$^{87}$Rb exhibits a finite positive scattering length of $100\,a_{0}$, while simultaneously varying the Rb-K scattering length on the negative side of the Feshbach resonance. The objective is to conduct comprehensive calculations involving cross-section, atom-dimer scattering length denoted by $a_{ad}$ for $^{40}$K-$^{87}$Rb$^{87}$Rb, and the energy of three-body bound states.
We employ basis sets with $N_{\theta}=168$ and $N_{\phi}=434$, ensuring the potential curves maintain at least six significant digits. Our computational setup utilizes approximately $10$ channels and $349$ sectors distributed across a range from $\rho = 18\, a_{0}$ to $\rho = 3100\, a_{0}$. Notably, the matching procedure in this section employs asymptotic wave functions expressed in rotated Jacobi coordinates, a technique known to enhance the convergence of observables concerning the matching distance $\rho_{m}$\,\cite{cyzhao2022}. In present calculations, we achieve the convergence of atom-dimer scattering lengths at $\rho_{m} = 2974\,a_{0}$. Fig.\,\ref{fig2a} illustrates the cross sections for elastic scattering of $^{40}$K + $^{87}$Rb$^{87}$Rb as a function of the $^{87}$Rb-$^{40}$K scattering length on the negative side of the Feshbach resonance. An evident peak is observed at approximately $a_{\scriptscriptstyle\textsl{RbK}}= - 155\,a_{0}$. We interpret the peak as being caused by the appearance of a three-body bound state at the atom-dimer threshold ($^{40}$K + $^{87}$Rb$^{87}$Rb). Owing to the presence of the near-threshold trimer state, the atom-dimer scattering length is expected to diverge. As illustrated in Fig.\,\ref{fig2b}, the $^{40}$K + $^{87}$Rb$^{87}$Rb elastic scattering lengths $a_{ad}$ are divergent at the same Rb-K scattering length value. A diagram of the relevant three-body energy spectrum as a function of Rb-K scattering length is shown in Fig.\,\ref{fig2c}, illustrating the energies of trimer states (the blue solid curve) and atom-dimer thresholds (the red dashed curve). The energy spectrum gives clear signatures at the threshold locations, which are consistent with the peak of the cross section as well as the divergence position of the atom-dimer scattering length. Hence, intriguing questions arise regarding whether the obtained three-body bound state, supported by the lowest hyperspherical potential curve, corresponds to an Efimov state and its connection to the observed atom-dimer resonance at positive Rb-K scattering lengths (or if it emerges from it). It is well-known that Efimov states possess universal properties that typically require scattering lengths significantly larger than van der Waals length $r_{\scriptscriptstyle\textsl{vdW}}$ and energies much smaller than the corresponding energy $E_{\scriptscriptstyle\textsl{vdW}}$. The Rb-Rb-K system can be characterized by a length $r_{\scriptscriptstyle\textsl{vdW,RbRb}}=83\,a_{0}$ and a corresponding energy $E_{\scriptscriptstyle\textsl{vdW,RbRb}}= 9.25\times 10^{-10}$ a.u. The three-body state is located at $|a_{\scriptscriptstyle\textsl{vdW,RbK}}|= 155\, a_{0}< 2\, r_{\scriptscriptstyle\textsl{vdW,RbRb}} $, with a binding energy larger than $E_{\scriptscriptstyle\textsl{vdW,RbRb}}$ by one order of magnitude. Therefore, the three-body state does not qualify as an Efimov state and the existence of two Efimov branches is ruled out.

	\subsubsection{Three-body recombination rates}
	We now consider three-body recombination, which is widely used to reveal Efimov physics in cold atoms. The event rate constant for three-body recombination is \\
	\begin{equation}
		\label{30}
		K_{3}^{(f\leftarrow i)}=\sum\limits_{J,\Pi}K_{3}^{J\Pi}=2!\sum\limits_{J,\Pi}\sum\limits_{f,i}\frac{32(2J+1)\pi^{2}}{\mu k^{4}}|S_{f \leftarrow i}^{J,\Pi}|^{2}\,.\\
	\end{equation}
	Here, $K_{3}^{J\Pi}$ is the partial recombination rate corresponding to the $J^{\Pi}$ symmetry, and $k=(2\mu E)^{1/2}$ is the hyperradial wavenumber in the incident channel. $S_{f \leftarrow i}^{J,\Pi}$ represents the scattering matrix element from the initial three-body continuum channels $i$ to the final atom-dimer channel $f$ for the $J^{\Pi}$ symmetry. The factor (2!) derives from the number of indistinguishable bosonic particles. The asymptotic form of the effective hyperradial potentials in $\rho \to \infty$ allows for generalization of Wigner's threshold law\,\cite{Esry2001Dec,Incao2005Jun} for recombination (i.e., the low-energy behavior of the recombination rates), which leads to the partial recombination rates $K_{3}^{J\Pi}$ satisfying\\
	\begin{equation}
		\label{31}
		K_{3}^{J\Pi}\propto E^{\lambda_{min}}\,.
	\end{equation}
	near threshold, where $\lambda_{min}$ is the minimum value of $\lambda$ in Eq.\,(\ref{16}).

\begin{figure}[htbp]
	\centering
	\subfigure{
		\includegraphics[width=7.8cm]{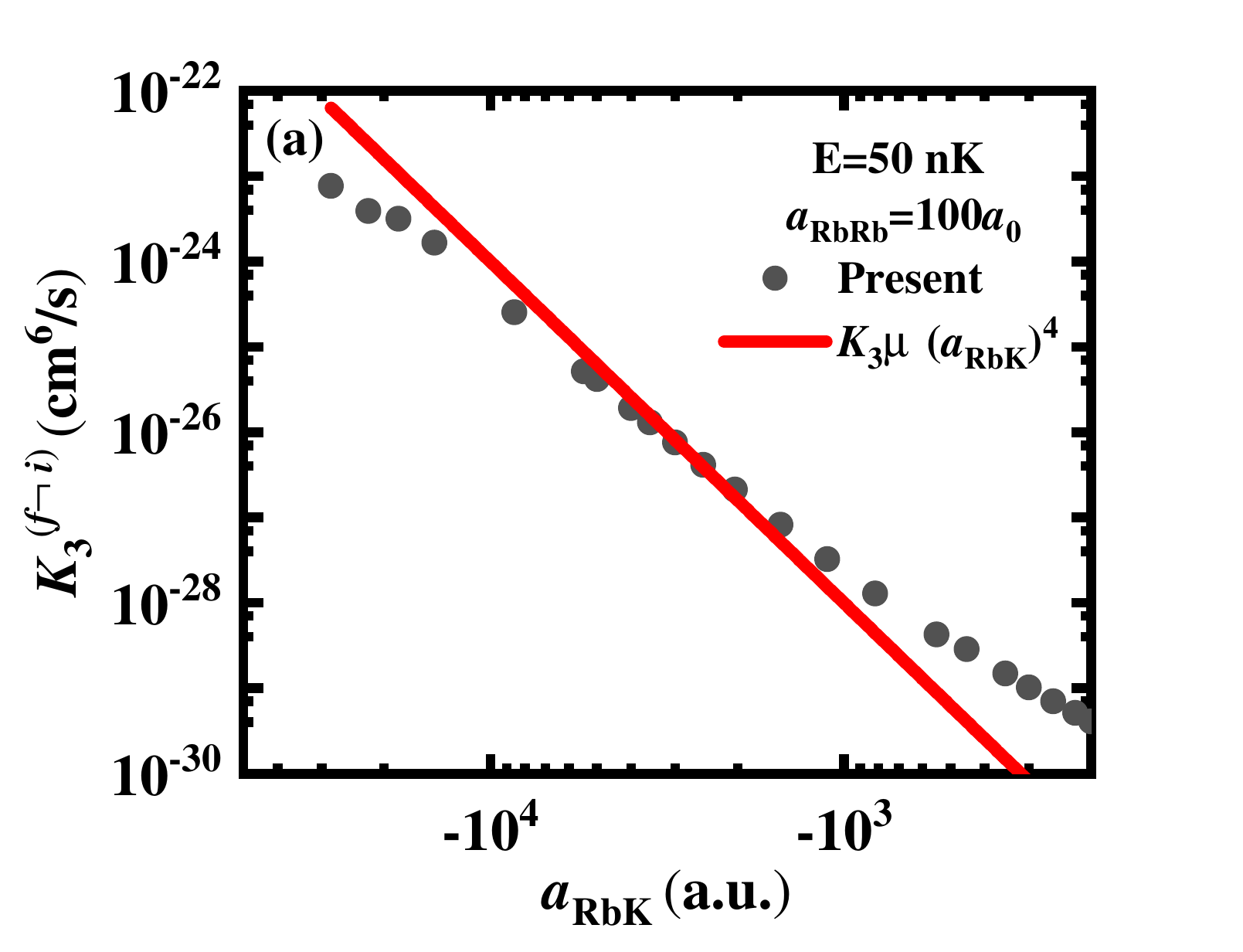}
		\label{fig3a}
	}
    \subfigure{
		\includegraphics[width=7.8cm]{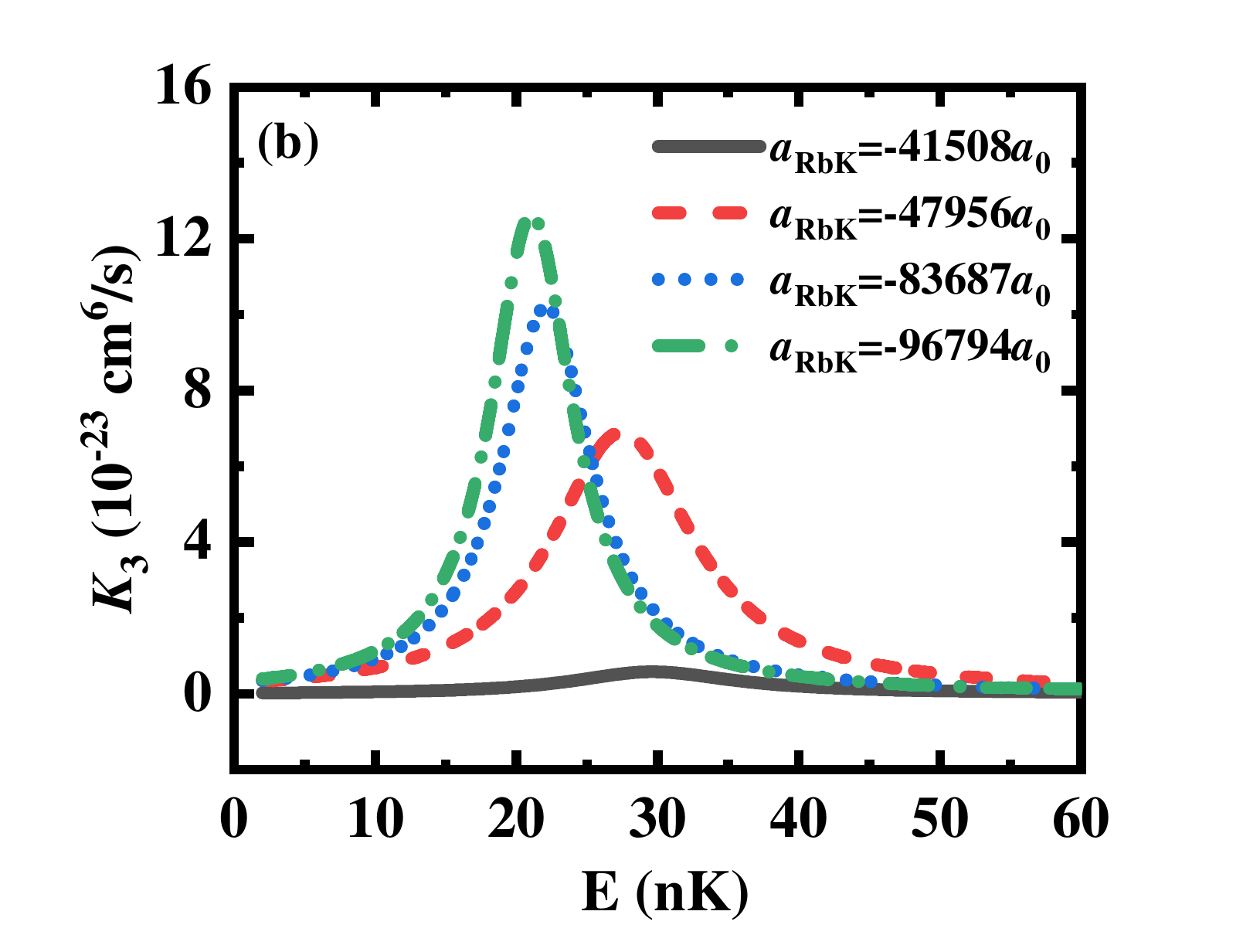}
		\label{fig3b}
	}
	\caption{(Color online) (a) The three-body recombination rates $K_{3}^{(f \leftarrow i)}$ for the $^{87}$Rb$^{87}$Rb$^{40}$K system as functions of $^{87}$Rb-$^{40}$K scattering lengths with $J^{\Pi}=0^{+}$ symmetry. (b) The three-body recombination rate as a function of collision energy
for several different negative scattering lengths near the resonance.}
	\label{fig3}
\end{figure}

\begin{figure}[htbp]
	\centering
	\subfigure{
		\includegraphics[width=7.8cm]{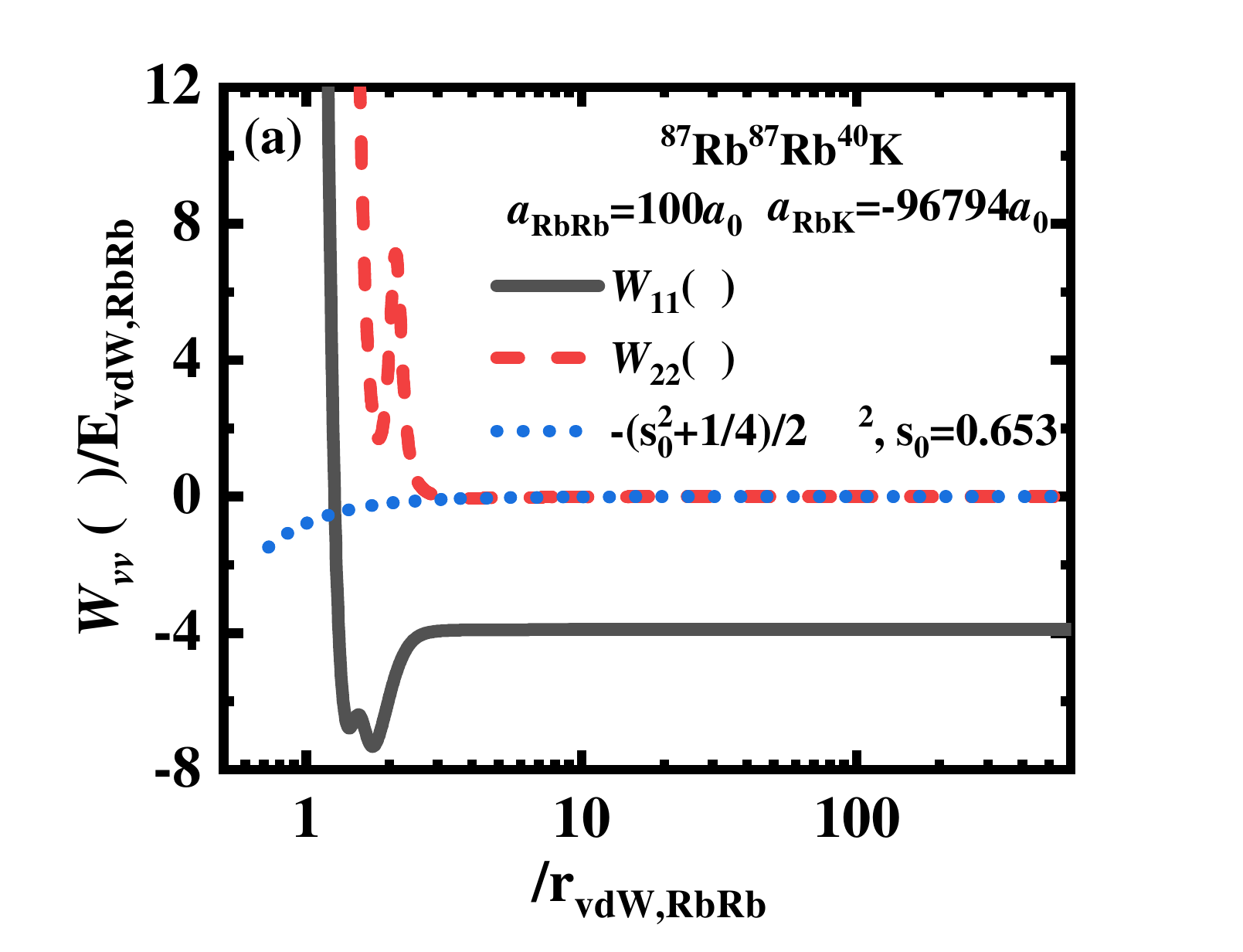}
		\label{fig4a}
	}
\subfigure{
		\includegraphics[width=7.8cm]{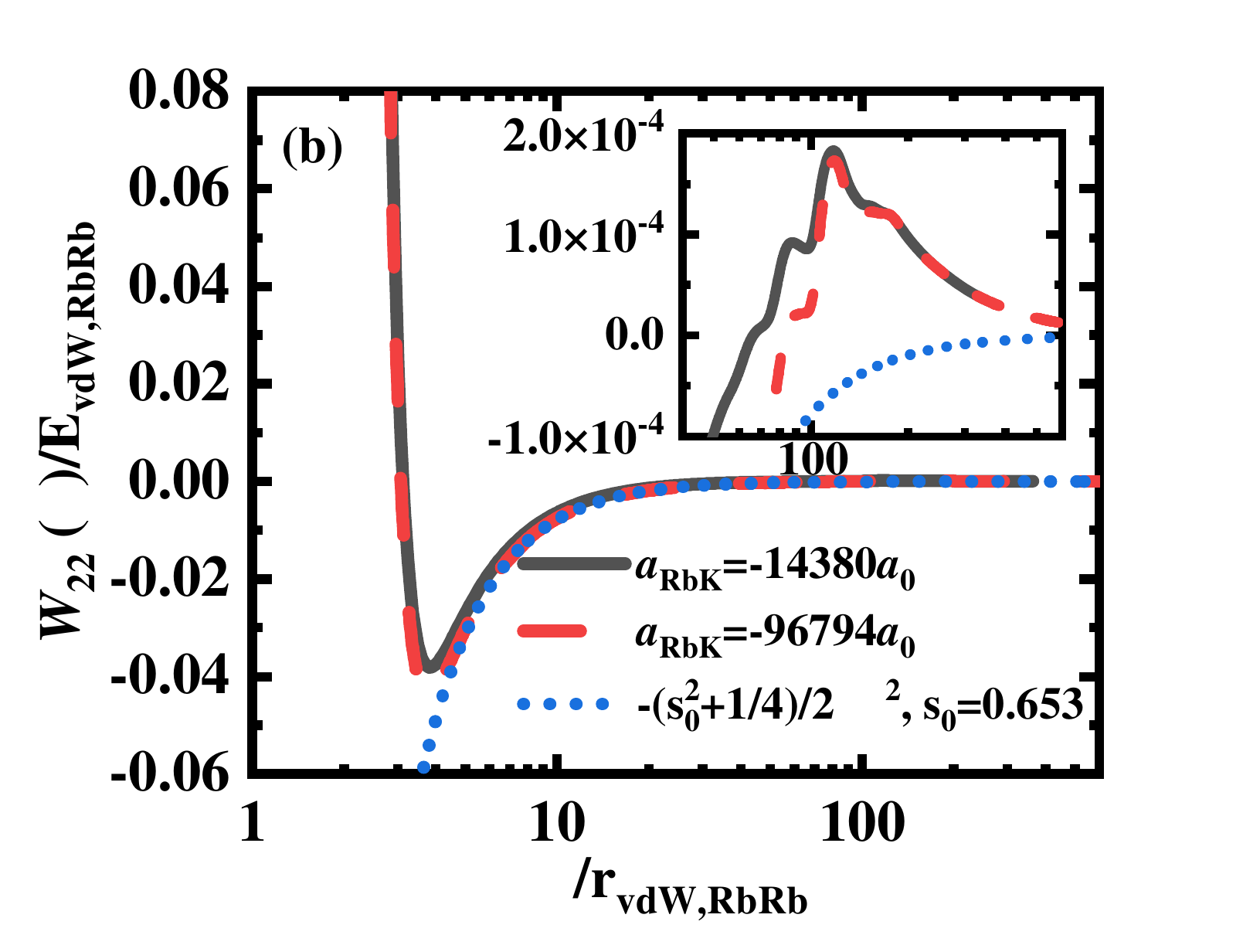}
		\label{fig4b}
	}
	\caption{(Color online) (a) The lowest two effective adiabatic potential curves of $^{87}$Rb$^{87}$Rb$^{40}$K, shown $a_{\scriptscriptstyle\textsl{RbK}} = -96794\,a_{0}$. The lowest potential curve approaches the atom-dimer threshold, while the second potential curve approaches the limit of three free atoms. (b) An enlarged view highlighting the potential barrier in the entrance channel.}
	\label{fig4}
\end{figure}

In Figure\,\ref{fig3a}, we present the numerically calculated three-body recombination rates, $K_{3}^{(f \leftarrow i)}$, for the $^{87}$Rb$^{87}$Rb$^{40}$K system at a collision energy of $E = 50$ nK. The solid line follows an $a^{4}$ dependence, and no Efimov features are seen for negative scattering lengths between $-200\,a_{0}$ and $-37000\,a_{0}$. Another approach to examining resonance evolution is to explore the energy-dependent behavior of recombination rates across different scattering lengths. Figure\,\ref{fig3b} shows these rates as a function of collision energy for several scattering lengths. Resonant behavior is evident here, with the peak position shifting closer to the three-body threshold as $|a_{\scriptscriptstyle\textsl{RbK}}|$ increases. The resonance becomes less pronounced for $a_{\scriptscriptstyle\textsl{RbK}} = - 41508\,a_{0}$. To further understand the connection between the recombination resonance in Fig.\,\ref{fig3b}, it is insightful to examine the effective adiabatic potentials in Fig.\,\ref{fig4}.

Figure\,\ref{fig4a} shows the lowest two effective adiabatic potential curves of the $^{87}$Rb$^{87}$Rb$^{40}$K system, specifically for $a_{\scriptscriptstyle\textsl{RbK}} = -96794 \,a_{0}$. The lowest channel represents the $^{40}$K + $^{87}$Rb$^{87}$Rb channel, which supports three-body bound states, while the other corresponds to the lowest entrance channel that approaches the Efimov curve at $\rho > 10\,r_{\scriptscriptstyle \textsl{vdW,RbRb}}$. Notably, there is a barrier in the lowest entrance channel that could support a three-body shape resonance. As $|a_{\scriptscriptstyle\textsl{RbK}}|$ increases, the barrier height decreases, as illustrated in the inset in Fig.\,\ref{fig4b}. This explains why the resonances position shifts toward the three-body threshold as $|a_{\scriptscriptstyle\textsl{RbK}}|$ increasing. The recombination rate is enhanced when the collision energy aligns with the resonance energy. Since the collision energy in an ultracold gas is determined by temperature, the interspecies scattering length can be varied to reveal resonances. As the scattering length changes, the resonance energy decreases with increasing $|a|$, eventually aligning with the collision energy at a particular $|a|$, thus yielding the observed resonance. If K-Rb mixtures can be cooled to around $600$ pK using specialized techniques, the useful range of $|a_{\scriptscriptstyle\textsl{RbK}}|$ can be extended to over $60000\,a_{0}$, allowing for the observation of these resonances\,\cite{Gong2019}.

\begin{figure}[htbp]
	\centering
	\subfigure{
		\includegraphics[width=7.8cm]{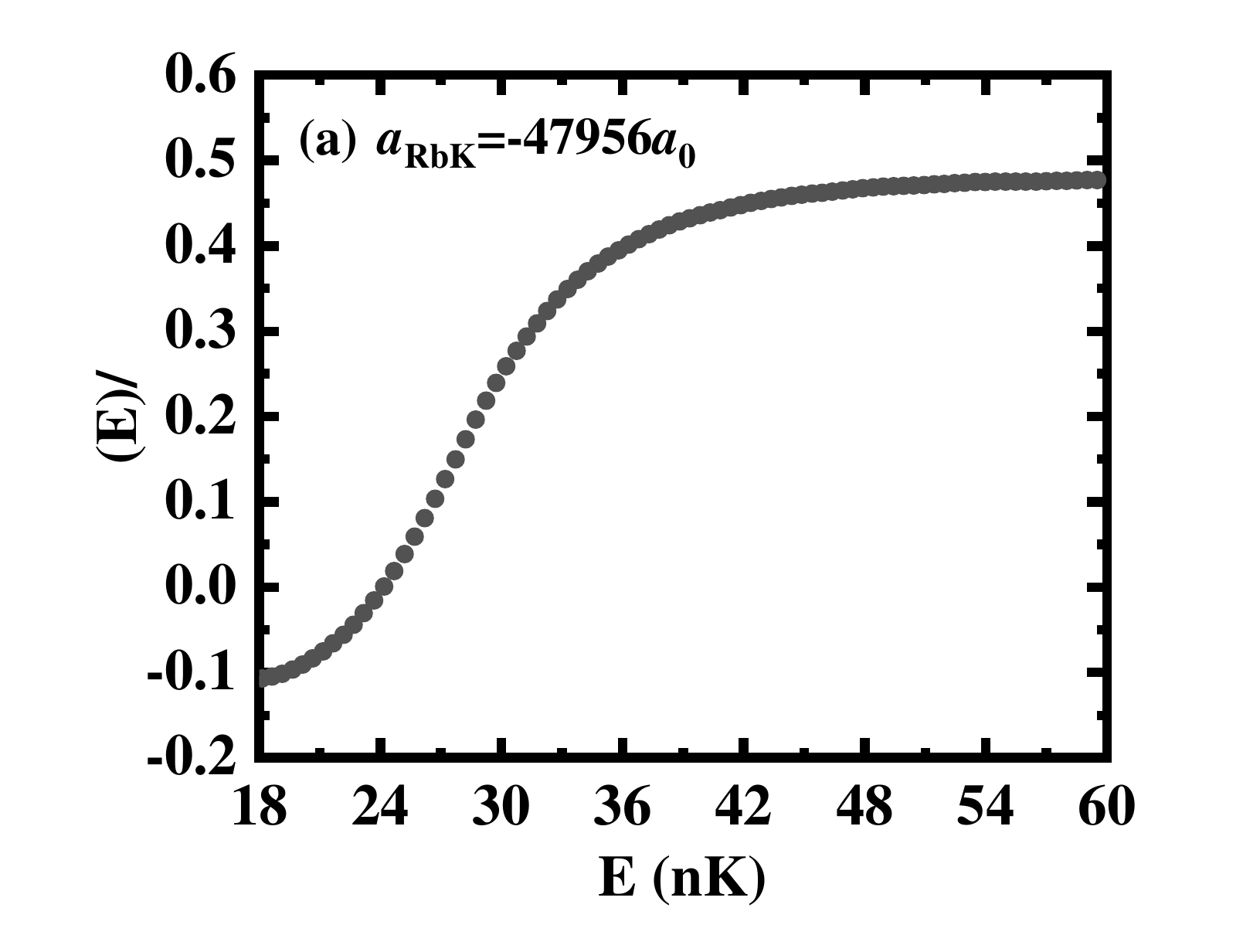}
		\label{fig54a}
	}
	\subfigure{
		\includegraphics[width=7.8cm]{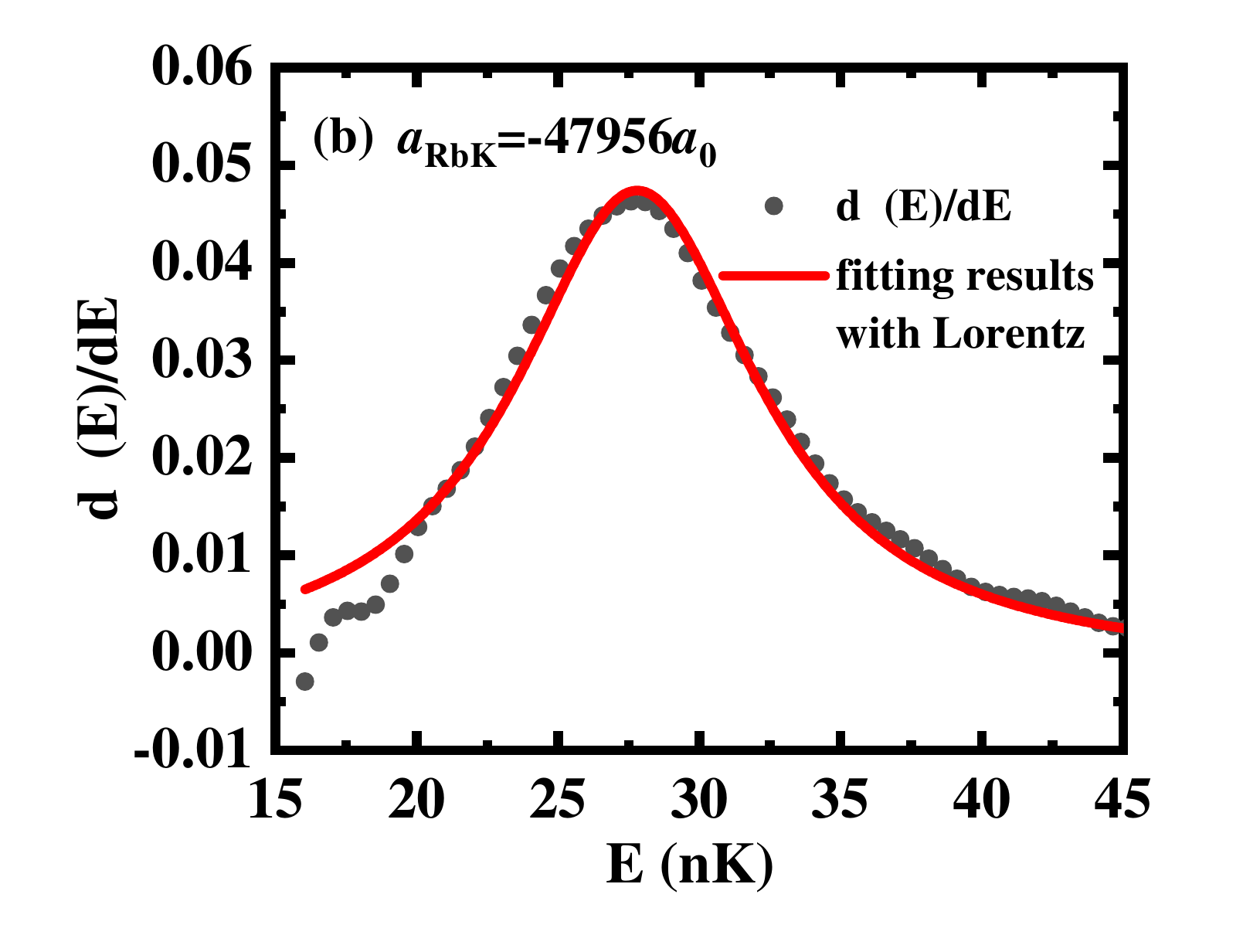}
		\label{fig5b}
	}
	\subfigure{
	\includegraphics[width=7.8cm]{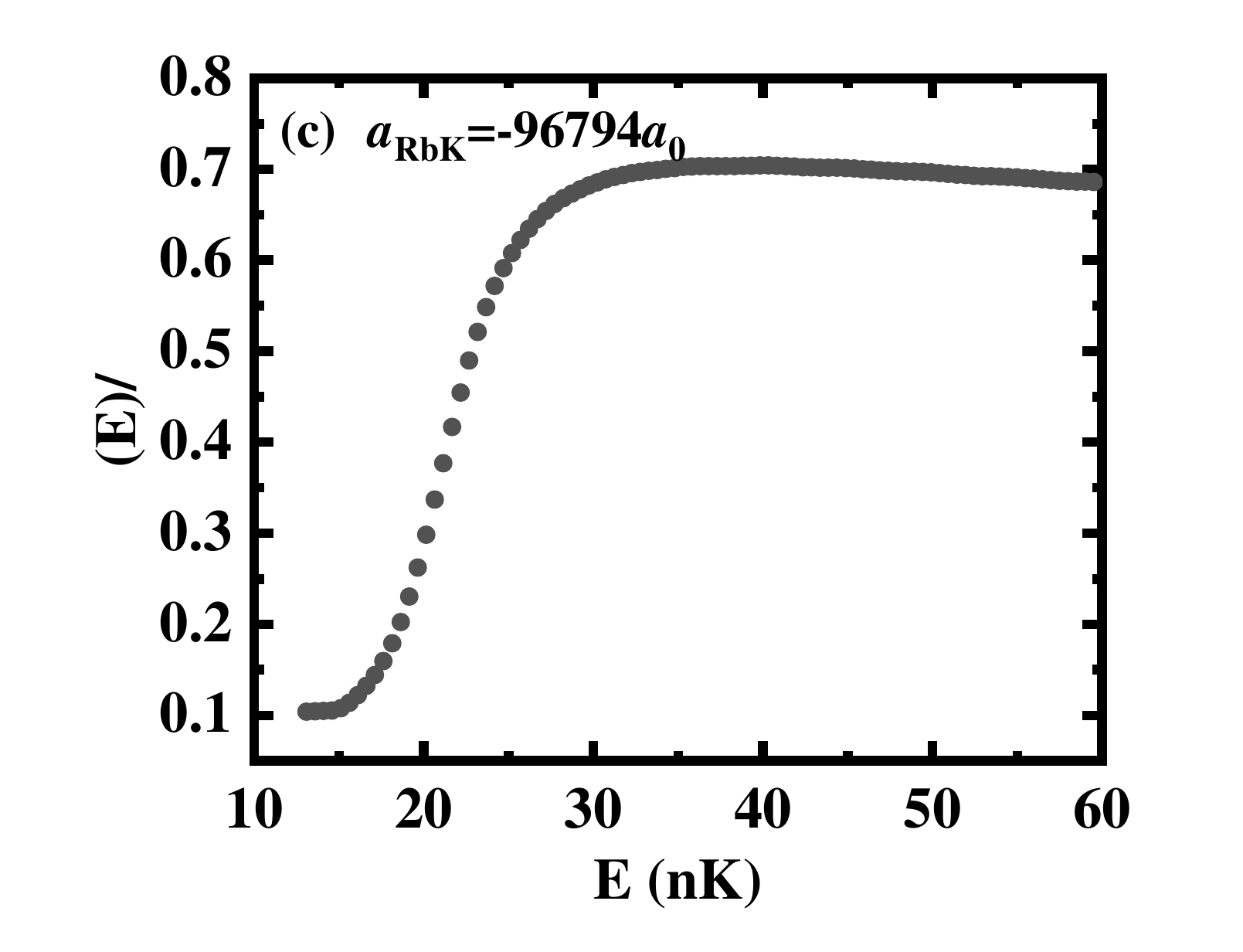}
	\label{fig5c}
}
\subfigure{
	\includegraphics[width=7.8cm]{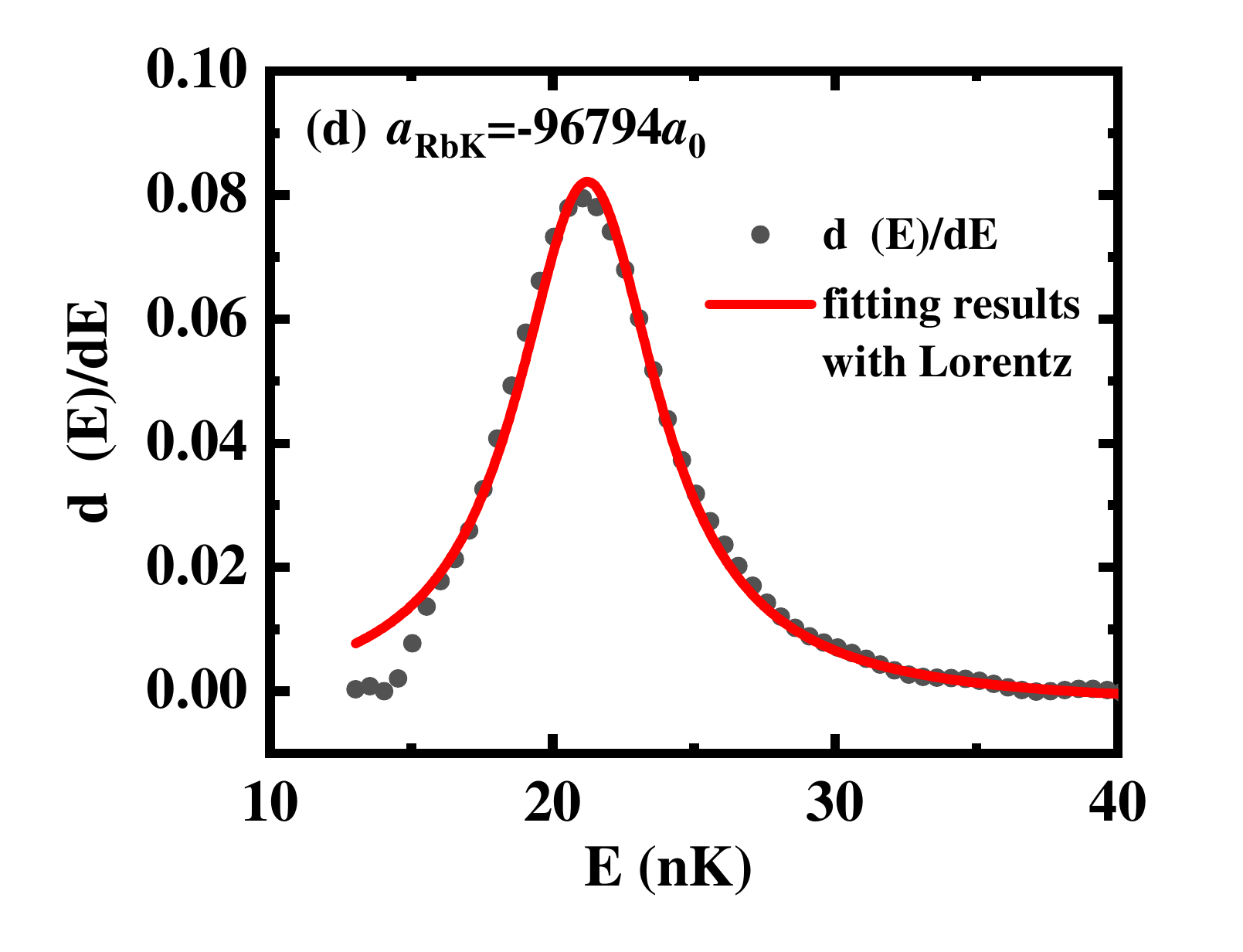}
	\label{fig5d}
}
	\caption{(Color online) (a) and (c) The corresponding eigenphase sum of $^{87}$Rb$^{87}$Rb$^{40}$K with  $a_{\scriptscriptstyle\textsl{RbRb}} =100\, a_0$, and $a_{\scriptscriptstyle\textsl{RbK}} = -47956\,a_{0}$ and $a_{\scriptscriptstyle\textsl{RbK}} = -96794\,a_{0}$, respectively. (b) and (d) Shown is the derivative of the sum of eigenphase shifts as a function of energy for $a_{\scriptscriptstyle\textsl{RbK}} = -47956\,a_{0}$ and $a_{\scriptscriptstyle\textsl{RbK}} = -96794\,a_{0}$, respectively. The red solid lines are the fitting results from the Lorentzian formula $d\delta (E)/dE=\delta_c+\frac{A}{\pi}\frac{\Gamma/2}{(E-E_{R})^2+(\Gamma/2)^2}$.}
\label{fig5}
\end{figure}

To analyze the three-body resonance energies $E_{R}$ and width $\Gamma$, we employ the sum of the eigenphase shifts (the eigenphase sum). The eigenphase shifts $\delta(E)$ are obtained by diagonalizing the $\mathcal{K}$-matrix followed by taking the arctan. Consequently, the total eigenphase shift is expressed as:\\
 	\begin{equation}
 	\label{32}
 	\delta_{tot}(E)=\sum\limits_{i=1}^{N_{o}}\delta_{i}(E)=\sum\limits_{i=1}^{N_{o}}arctan(\lambda_{i})\,.
 \end{equation}
  where $\lambda_{i}$ is the $i$-th eigenvalue of $\mathcal{K}$-matrix, $E$ is the collision energy, and $N_o$ is the number of open channels. The resonant position is the point at which, the time delay $\tau = \frac{d \delta_{tot}(E)}{dE}$ is maximal. Fig.\,\ref{fig5} shows the eigenphase sum in Eq.\,(\ref{32}) and the energy derivative of the eigenphase sum. The resonance peak corresponds to the three-body shape resonances, with $E_{R} = 21.2$ nK, $\Gamma=6.11$ for $a_{\scriptscriptstyle\textsl{RbK}} = -96794\,a_{0}$, and $E_{R}=27.8$ nK, $\Gamma=10.6$ for $a_{\scriptscriptstyle\textsl{RbK}} = -47956\, a_{0}$, respectively.

\subsection{On the positive side of the $^{87}$Rb-$^{40}$K scattering length}

\subsubsection{Three-body recombination rates}
\begin{figure}[htbp]
		\centering
		\subfigure{
			\includegraphics[width=7.8cm]{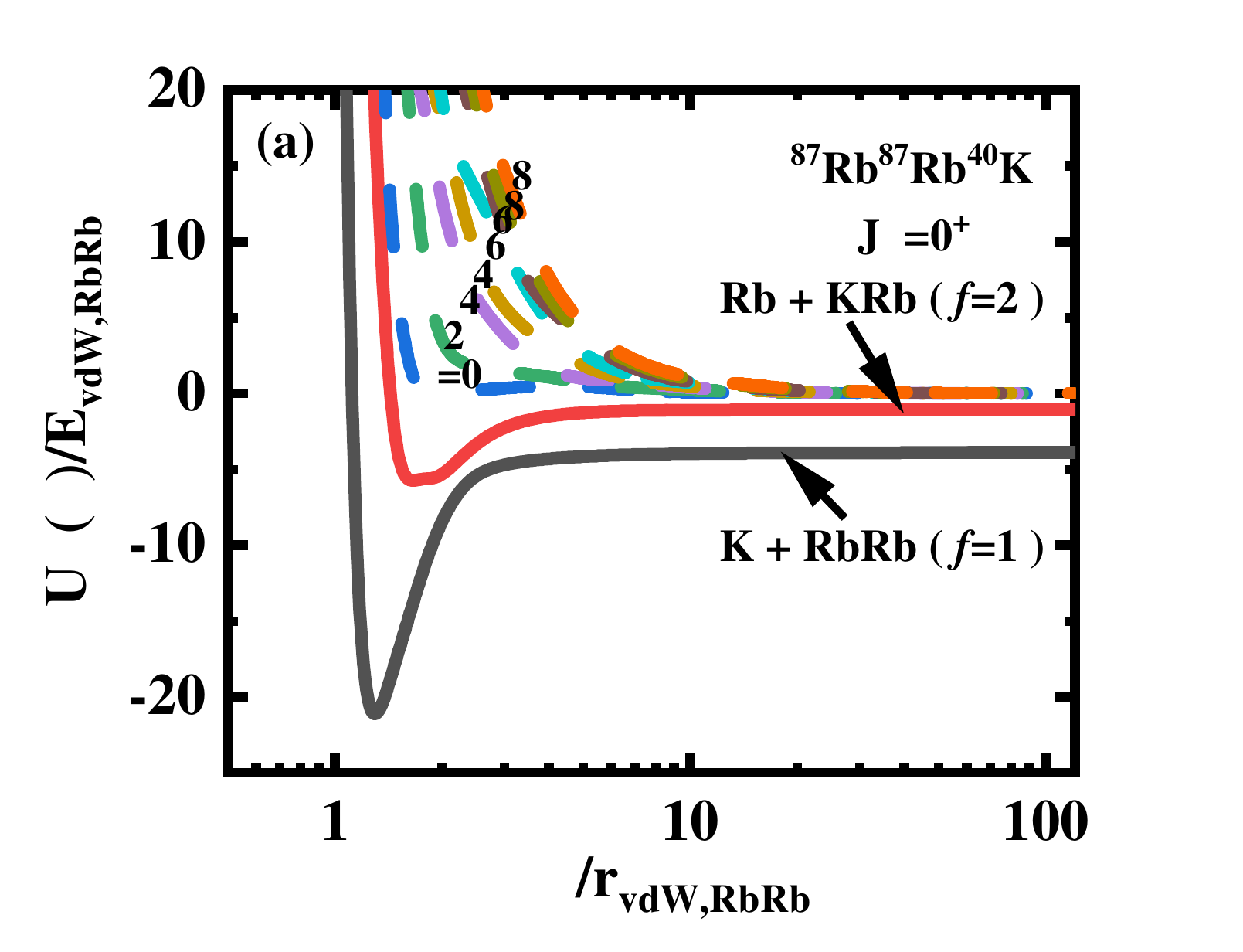}
			\label{fig6a}
		}
		\subfigure{
			\includegraphics[width=7.8cm]{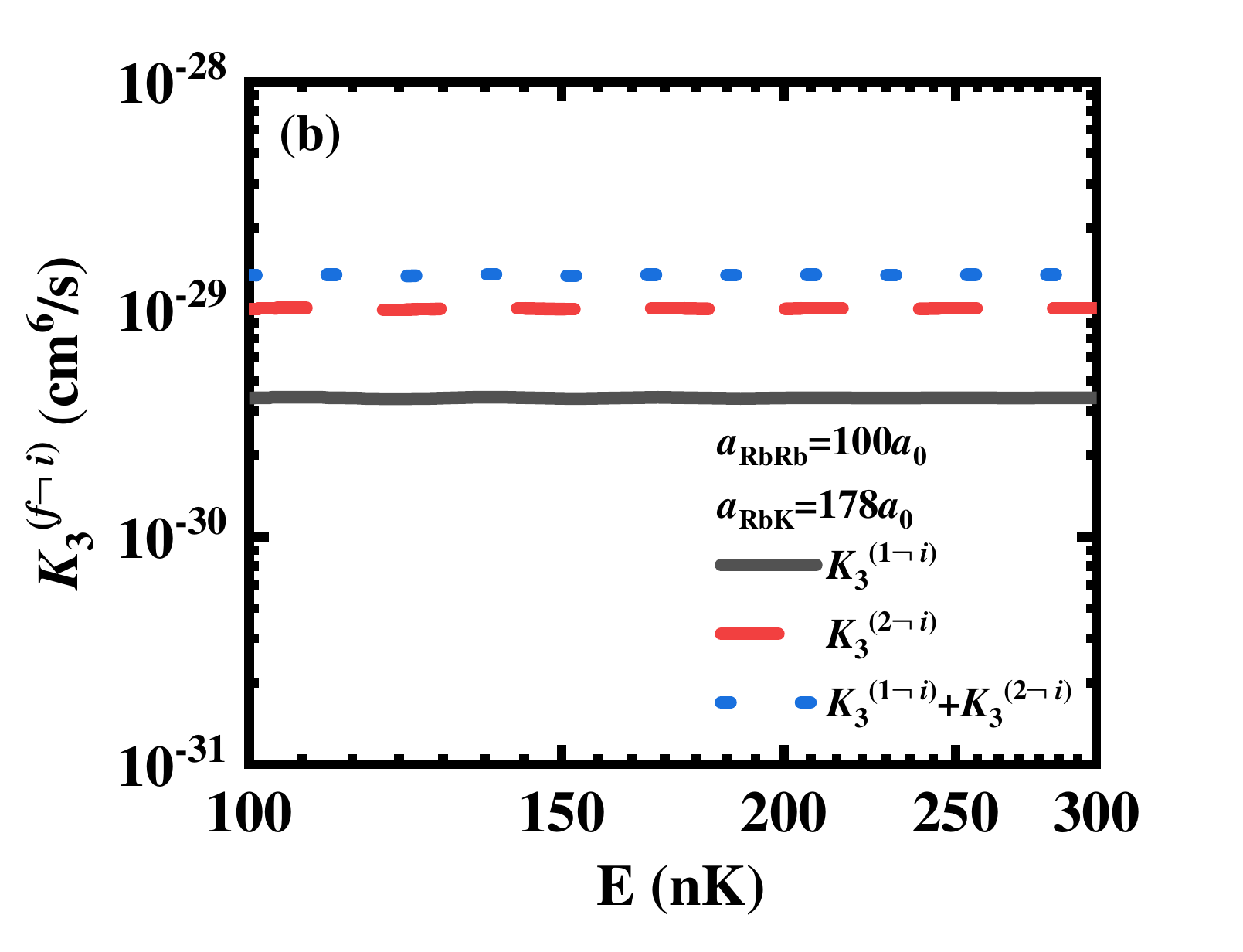}
			\label{fig6b}
		}
		\caption{(Color online) (a) The $J^{\Pi}=0^{+}$ adiabatic hyperspherical potential curves U$_{\nu}$($\rho$) for $^{87}$Rb-$^{87}$Rb-$^{40}$K with $a_{\scriptscriptstyle\textsl{RbRb}}=100\,a_{0}$ and $a_{\scriptscriptstyle\textsl{RbK}}=178\,a_{0}$. The values of $\lambda$ indicate the asymptotic behavior of the potential curves representing the three-body continuum channels, as given in Eq.\,(\ref{16}). The lowest adiabatic potential curve denotes a recombination channel that corresponds to a K atom and an RbRb dimer. (b) The three-body recombination rate $K_{3}^{(f \leftarrow i)}$ as a function of the collision energy $E$ for the $^{87}$Rb$^{87}$Rb$^{40}$K system in the $J^{\Pi}=0^{+}$ symmetry with $a_{\scriptscriptstyle\textsl{RbRb}}=100\,a_{0}$ and $a_{\scriptscriptstyle\textsl{RbK}}=178\,a_{0}$. The black solid line and red dashed line indicate three-body recombination rates for the sum of different incident channels $i$ to particular recombination channels $f=1,2$, respectively. The blue dotted line indicates the sum of the above two three-body recombination rates. The recombination rates $K_{3}^{(f \leftarrow i)}$ follow the Wigner threshold law prediction, $K_{3}^{(f \leftarrow i)}\propto E^{0}$ in the low-energy region.}
		\label{fig6}
	\end{figure}

	In this part, we first study the three-body recombination rates $K_{3}^{(f\leftarrow i)}$ for the $^{87}$Rb$^{87}$Rb$^{40}$K system with $J^{\Pi}=0^{+}$ symmetry on the positive Rb-K scattering length side. Figure\,\ref{fig6a} shows the hyperspherical potential curves for the $^{87}$Rb-$^{87}$Rb-$^{40}$K system, in which the Rb-K scattering length is $a_{\scriptscriptstyle\textsl{RbK}}=178\,a_{0}$ and Rb-Rb interacts via the s-wave scattering length $a_{\scriptscriptstyle\textsl{RbRb}}=100\,a_{0}$. It is shown that there are two recombination channels: one being the weakly bound $^{87}$Rb + $^{40}$K$^{87}$Rb channel labeled as $f=2$ and the other being the deeply bound channel $^{40}$K + $^{87}$Rb$^{87}$Rb labeled as $f=1$. Figure\,\ref{fig6b} shows the three-body recombination rate $K_{3}^{(f\leftarrow i)}$ as a function of the collision energy $E$ for the $^{87}$Rb$^{87}$Rb$^{40}$K system with $J^{\Pi}=0^{+}$ symmetry. At lower collision energies, the recombination rates $K_{3}^{(f\leftarrow i)}$ are constant and follow the Wigner threshold law prediction, $K_{3}^{(f \leftarrow i)}\propto E^{0}$.\\
	
\begin{figure}[htbp]
		\centering
		\subfigure{
			\includegraphics[width=7.8cm]{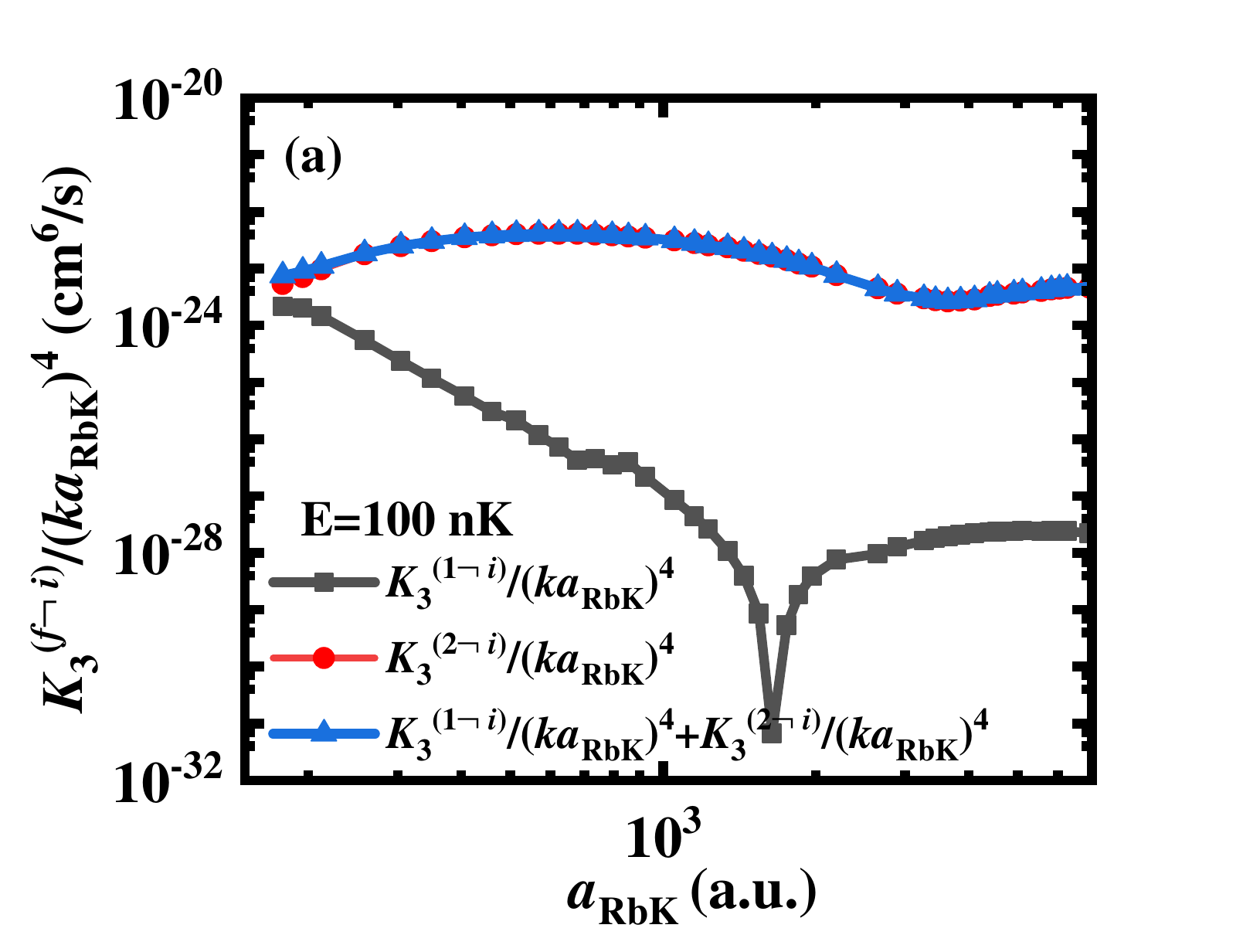}
			\label{fig7a}
		}
		\subfigure{
			\includegraphics[width=7.8cm]{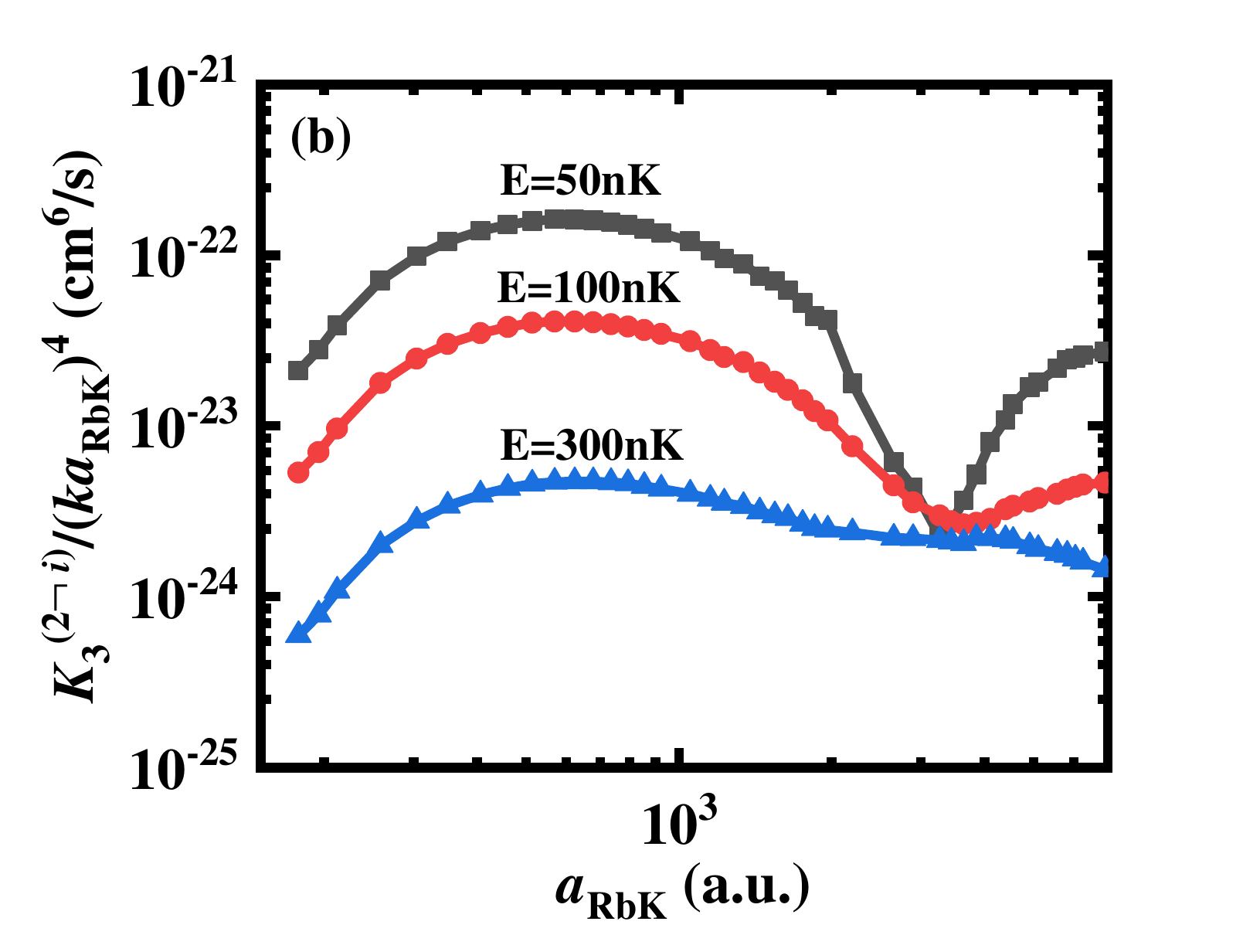}
			\label{fig7b}
		}
		\caption{(Color online) (a) The $J^{\Pi} = 0^{+}$ three-body recombination rate $K_{3}^{(f \leftarrow i)}$ is shown as a function of Rb-K scattering length $a_{\scriptscriptstyle\textsl{RbK}}$ for $E=100$ nK. The Rb-Rb interaction is fixed at $a_{\scriptscriptstyle\textsl{RbRb}}=100\,a_{0}$. The filled squares (black) represent recombination of the deep channel Rb + Rb+ K $\to$ K + RbRb, the filled circles (red) represent recombination of the shallow channel Rb + Rb + K $\to$ Rb + KRb, and the filled triangles (blue) depict the sum of the channels. The local minimum is at $a_{\scriptscriptstyle\textsl{RbK}}=1641\,a_{0}$ in the deep recombination channel. An Efimov recombination minimum on the total rates is shown at $a_{\scriptscriptstyle\textsl{RbK}} = 3638\,a_{0}$. (b) Magnifying the position of Efimov recombination minimum for different collision energies.}
		\label{fig7}
	\end{figure}

	Figure\,\ref{fig7a} shows the $a_{\scriptscriptstyle\textsl{RbK}}$ dependence of the three-body recombination rates at positive Rb-K scattering lengths with the Rb-Rb interaction fixed at $a_{\scriptscriptstyle\textsl{RbRb}}=100\,a_{0}$. The data are obtained at a collision energy of $E=100$ nK. The filled black squares represent recombination into the deeply bound channel K + RbRb, red solid circles represent recombination into the weakly bound channel Rb + KRb, and blue solid triangles depict their sums. The recombination rates of the deep channel (K + RbRb) are close to those of the shallow channel (Rb + KRb) at small Rb-K scattering lengths. Then, the recombination rates of the deep channel begin to decrease when the Rb-K scattering lengths are tuned larger. A minimum appears at approximately $a_{\scriptscriptstyle\textsl{RbK}}=1641\,a_{0}$ in the deep recombination channel. However, this recombination minimum has no effect on the total recombination rates since it is smaller than that into the weakly bound channel by at least 3 orders of magnitude. An Efimov recombination minimum for total rates arises at approximately $a_{\scriptscriptstyle\textsl{RbK}} = 3638\,a_{0}$. For $a > 0$, the three-body recombination minimum is a well-known feature of Efimov physics\,\cite{Zaccanti2009,Pollack2009}, which is explained as the destructive interference between two different decay pathways.
	Figure\,\ref{fig7b} shows the dependence of Efimov recombination minimum on the collision energy. The recombination minimum is at $a_{\scriptscriptstyle+} = 3264\,a_{0}$ with the collision energy of $E = 50$ nK and is $ a_{\scriptscriptstyle+} = 3638\,a_{0} $ for $E = 100$ nK. However, this minimum is less evident at the collision energy of $E = 300$ nK.
	
	\subsubsection{Atom($^{87}$Rb)-dimer($^{40}$K$^{87}$Rb) inelastic scattering }
\begin{figure}[htbp]
		\centering
		\subfigure{
			\includegraphics[width=7.8cm]{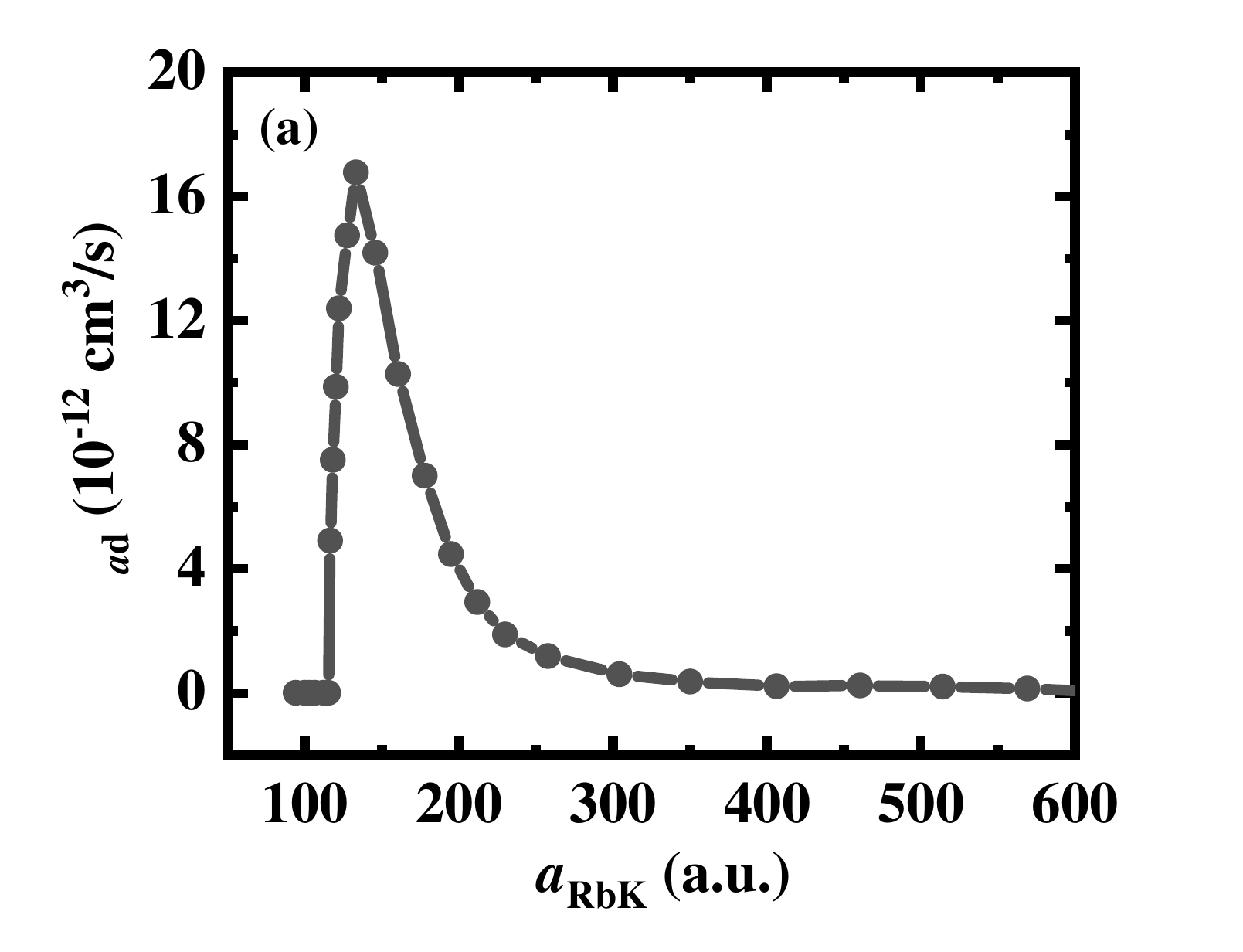}
			\label{fig8a}
		}
		\subfigure{
			\includegraphics[width=7.8cm]{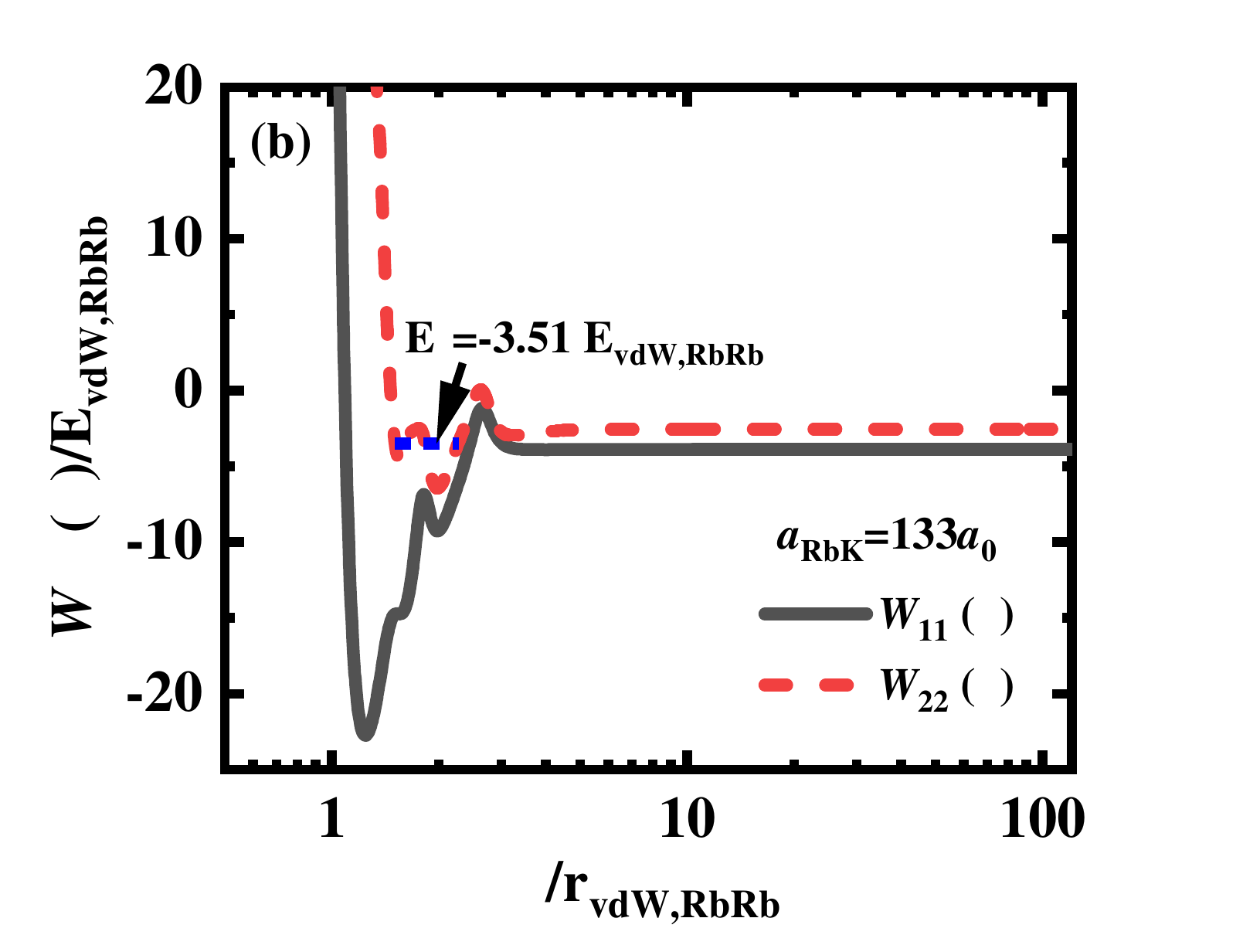}
			\label{fig8b}
		}
		\caption{(Color online) (a) The $J^{\Pi} = 0^{+}$ atom-dimer loss coefficients in the $^{87}$Rb-$^{87}$Rb-$^{40}$K system, with the Rb-Rb interaction set at $a_{\scriptscriptstyle\textsl{RbRb}} = 100\,a_0$, showing a resonant feature at $a_{\scriptscriptstyle\textsl{RbK}} = 133\,a_0$. The collision energy is $E =0.001$ nK. (b) The lowest two effective adiabatic potential curves $W_{\nu\nu}(\rho)$ for $^{87}$Rb-$^{87}$Rb-$^{40}$K system with $a_{\scriptscriptstyle\textsl{RbRb}} = 100\,a_0$ and $a_{\scriptscriptstyle\textsl{RbK}} = 133\,a_0$. The position of the resonant state is labeled on the figure (blue dotted line). }
		\label{fig8}
	\end{figure}

On the positive side of the Feshbach resonance ($a > 0$), a key observable is the atom-dimer loss coefficient, $\beta_{ad}(a)$. This coefficient relates to the inelastic atom-dimer scattering cross-section through $\beta_{ad}(a) = \frac{k_{ad}}{\mu_{ad}} \sigma_{ad}^{\text{inel}}$, where $k_{ad} = \sqrt{2 \mu_{ad}(E-E_{2b})}$, $E_{2b}$ is the dimer binding energy, and $\mu_{ad}$ is the atom-dimer reduced mass. Figure\,\ref{fig8a} presents the calculated atom-dimer loss coefficient $\beta_{ad}(a)$ for the $^{87}$Rb-$^{87}$Rb-$^{40}$K system as a function of the $^{87}$Rb-$^{40}$K scattering length. A resonant feature is observed starting at $a_{\scriptscriptstyle\textsl{RbK}} \geq 133\,a_0$, attributed to a three-body resonant state coinciding with the atom-molecule threshold. Our single-channel model does not reproduce the experimental results well, indicating the significant impact of multichannel Feshbach physics on the systems.

\begin{figure}[htbp]
	\centering
	\subfigure{
		\includegraphics[width=7.8cm]{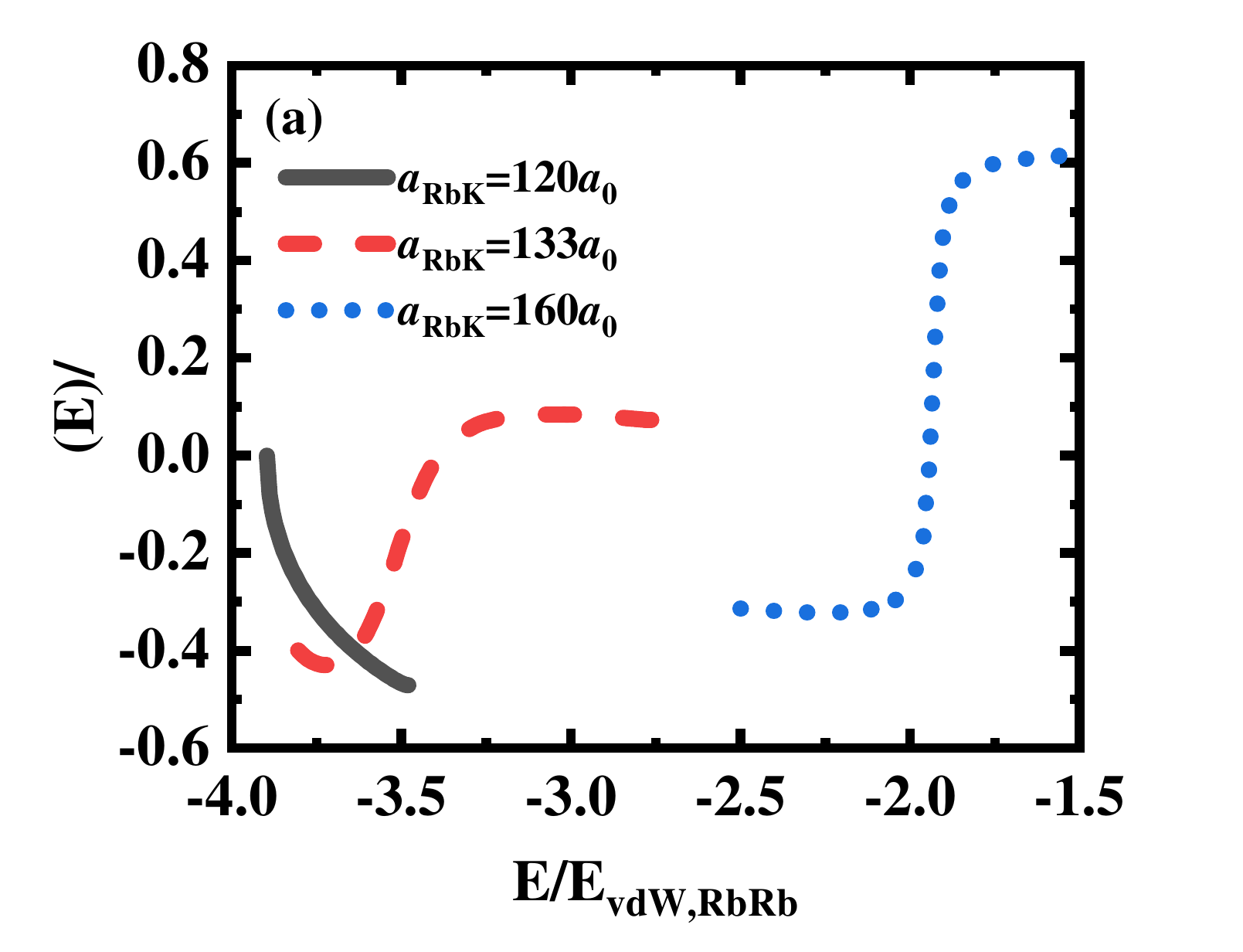}
		\label{fig9a}
	}
\subfigure{
		\includegraphics[width=7.8cm]{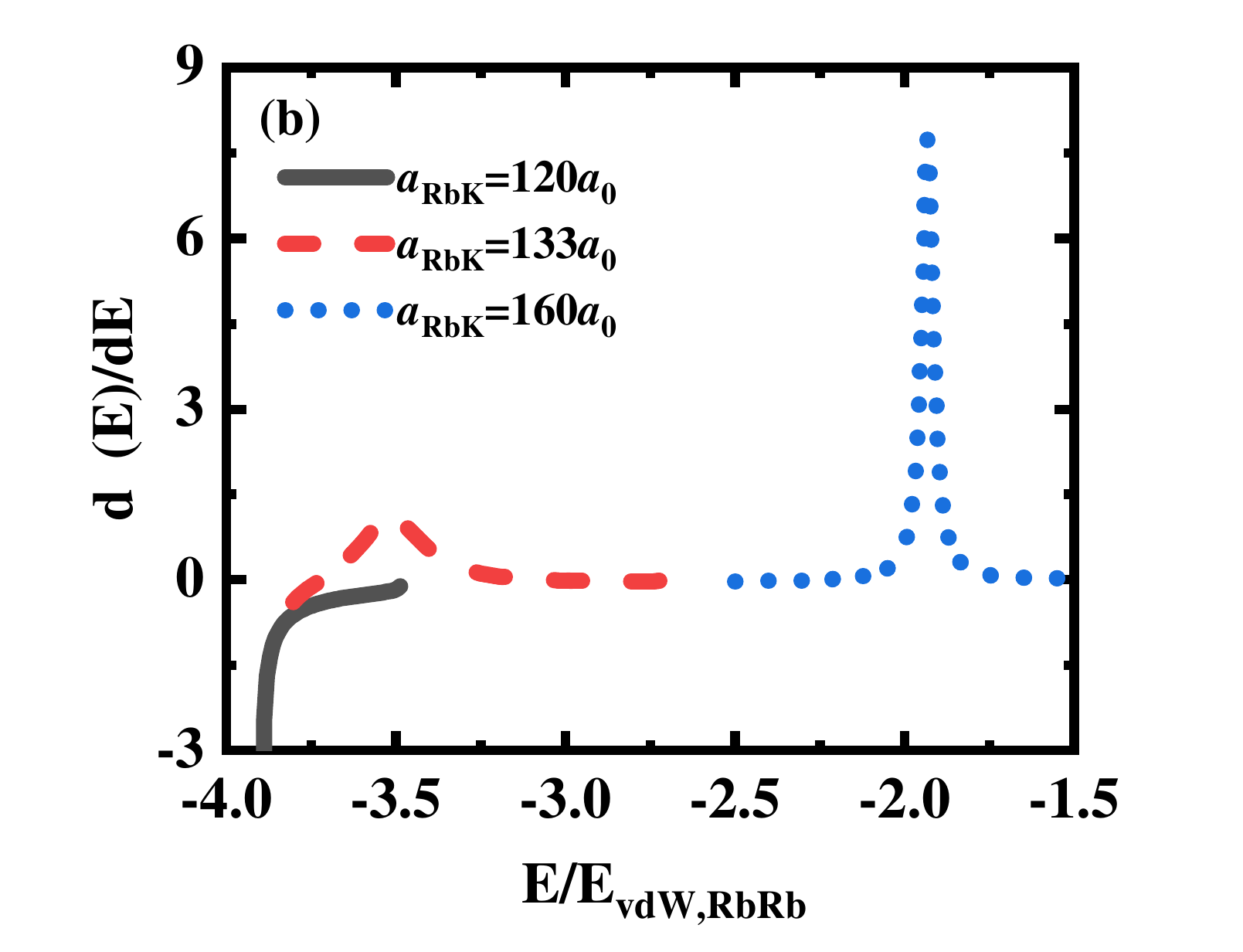}
		\label{fig9b}
	}
	\caption{(Color online) (a) The eigenphase sum for the $^{87}$Rb-$^{87}$Rb-$^{40}$K system with $a_{\scriptscriptstyle\textsl{RbK}} = 120,\,133,\,160\,a_0$ and the Rb-Rb interaction fixed at $a_{\scriptscriptstyle\textsl{RbRb}} = 100\,a_0$. (b) The corresponding energy derivative of the eigenphase sum, $d\delta(E)/dE$.}
	\label{fig9}
\end{figure}

The energy of this resonant state is obtained using the eigenphase sum method. Figure\,\ref{fig9} displays the eigenphase sum in Eq.\,(\ref{32}) and its energy derivative at various scattering lengths: $a_{\scriptscriptstyle\textsl{RbK}} = 120,\,133,\,160\,a_0$. The results show a resonant state energy of $E_{R} = -3.51\,E_{\scriptscriptstyle\textsl{vdw,RbRb}}$, $\Gamma=0.215$ for $a_{\scriptscriptstyle\textsl{RbK}} = 133\,a_0$, and $E_{R} = -1.93\,E_{\scriptscriptstyle\textsl{vdw,RbRb}}$, $\Gamma=0.041$ for $a_{\scriptscriptstyle\textsl{RbK}} = 160\,a_0$. No resonant state is observed at $a_{\scriptscriptstyle\textsl{RbK}} = 120\,a_0$. Figure\,\ref{fig8b} illustrates the effective potential curves for the $^{87}$Rb-$^{87}$Rb-$^{40}$K system, with $a_{\scriptscriptstyle\textsl{RbK}} = 133\,a_0$ and $a_{\scriptscriptstyle\textsl{RbRb}} = 100\,a_0$. This figure shows that the resonant state is associated with the Rb + KRb channel, with the position of the resonant state labeled.

\begin{figure}[htbp]
		\centering
		\subfigure{
			\includegraphics[width=7.8cm]{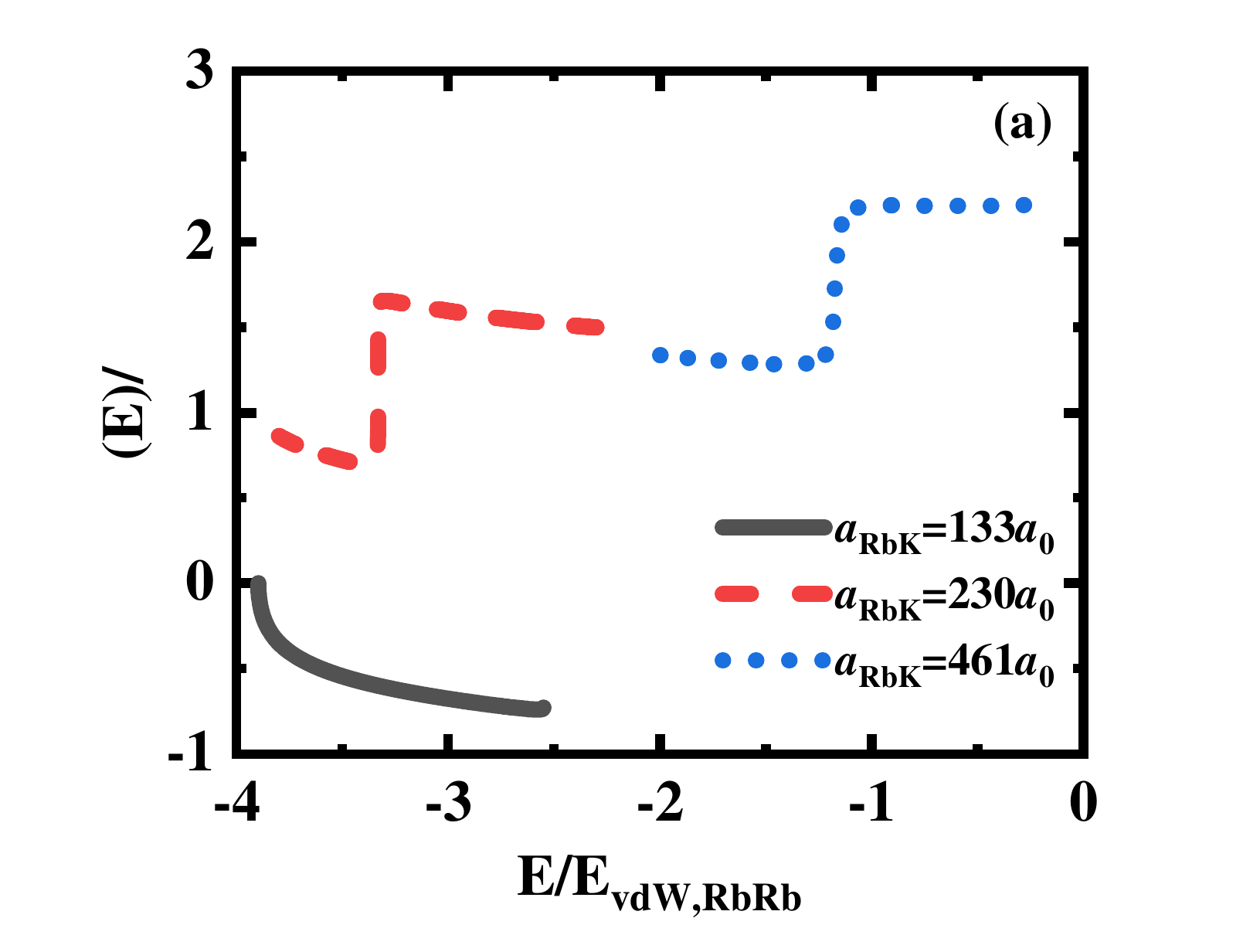}
			\label{fig10a}
		}
			\subfigure{
		\includegraphics[width=7.8cm]{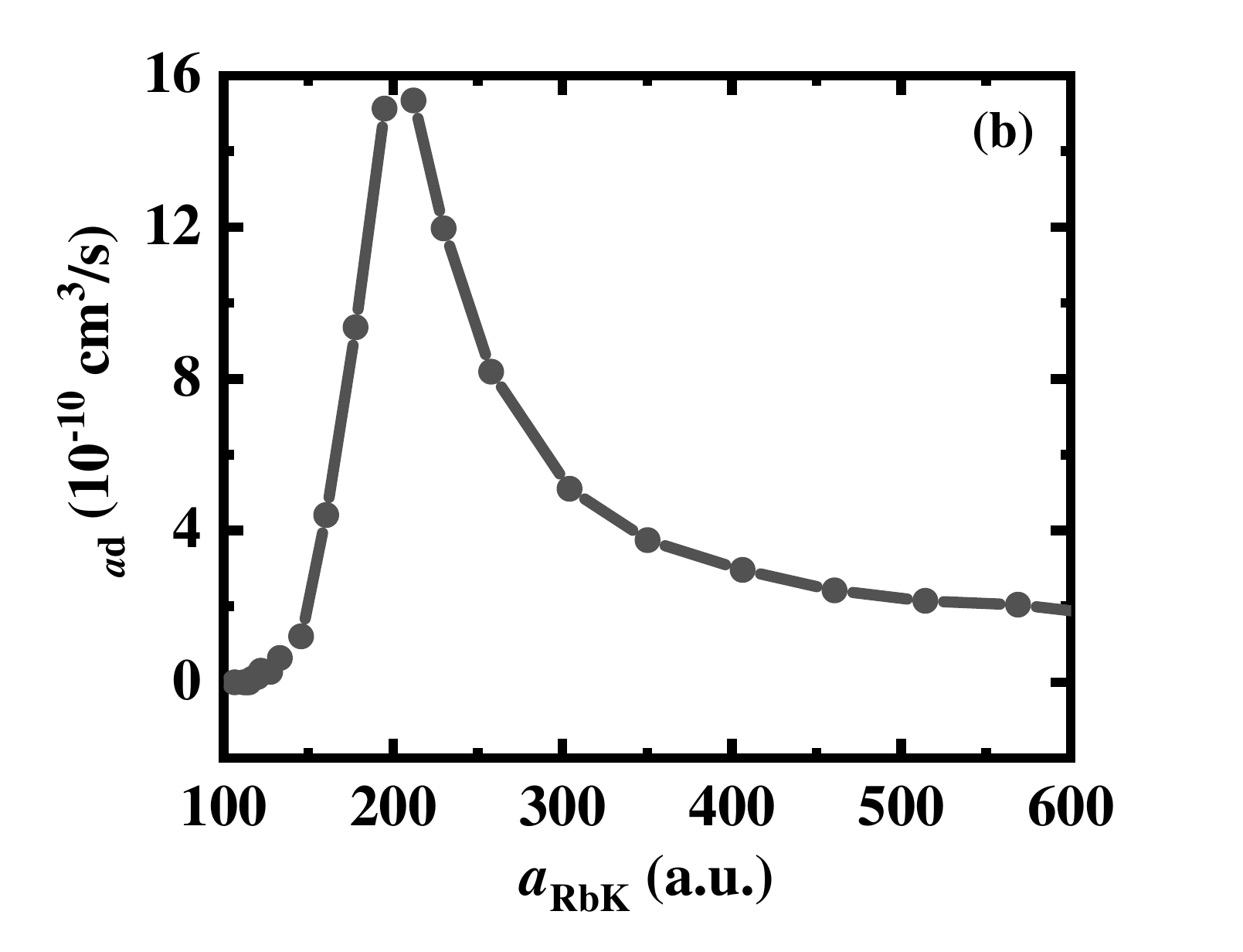}
		\label{fig10b}
	}
		\caption{(Color online) With the inclusion of a three-body interaction: (a) The eigenphase sum for $a_{\scriptscriptstyle\textsl{RbK}} = 133,\,230,\,461 \,a_0$. (b) The atom-dimer loss coefficients in the $^{87}$Rb-$^{87}$Rb-$^{40}$K system.}
		\label{fig10}
	\end{figure}

To align our model with experimental observations, we introduce a three-body interaction of the form \( V(\rho) = -A \rho^B e^{-\rho / \beta} \), with parameters \( B = 2 \) and \( \beta = 0.2\, r_{\scriptscriptstyle\textsl{vdW,RbRb}} \)\,\cite{Yudkin2024}. When the parameter \( A \) is tuned to \( A = 5.6 \), the atom-dimer resonance position closely matches the experimental value. Figure \ref{fig10} shows the calculated atom-dimer loss coefficient \( \beta_{ad}(a) \) and eigenphase sum for the $^{87}$Rb-$^{87}$Rb-$^{40}$K system as a function of the $^{87}$Rb-$^{40}$K scattering length, incorporating the three-body interaction. A resonant feature is observed at \( a_{\scriptscriptstyle\textsl{RbK}} \geq 212\,a_0 \).

\subsection{Discussion of the universal relationship between the different Efimov features }
The universal relations among various Efimov features were first predicted in the zero-range limit by Helfrich et al.\,\cite{Helfrich2010}, utilizing an effective-field-theory framework. According to the zero-range predictions, the universal relationship for the locations of Efimov features associated with the same trimer state is $|a_{\scriptscriptstyle-}|/a_{\scriptscriptstyle*}=240$\,\cite{D_Incao_2018}. Given an observed resonance at $a_{\scriptscriptstyle*} = 230\,a_0$\,\cite{Zirbel2008,Bloom2013}, the zero-range limit would imply a value of $a_{\scriptscriptstyle-} = -55200\,a_0$.

In the study by Acharya et al.\,\cite{Acharya2016}, a more comprehensive set of universal relations was derived to connect different Efimov features, accounting for finite interspecies effective range and intraspecies scattering length effects, as outlined in Eq. (25) of their paper. Applying this equation requires two Efimov features as inputs. For the $^{87}$Rb-$^{87}$Rb-$^{40}$K system under investigation, we obtained a numerical value of $a_{\scriptscriptstyle+} = 3264\,a_0$ at a collision energy of $E=50$ nK. Using this value and incorporating the parameters $a_{\scriptscriptstyle*} = 230\,a_0$, $a_{22} = a_{\scriptscriptstyle\textsl{RbRb}} = 100\,a_0$, and $r_{0} = r_{\scriptscriptstyle \textsl{vdW,RbK}} = 72\,a_0$, Eq. (25) from Ref.\,\cite{Acharya2016} predicts the first Efimov resonance to occur at $a_{\scriptscriptstyle-} = -48304\,a_0$.

Using the universal relation $a_{\scriptscriptstyle*}/a_{\scriptscriptstyle+} = 0.046$ and the observed resonance at $a_{\scriptscriptstyle*} = 230\,a_0$, we find $a_{\scriptscriptstyle+} = 5020\,a_0$. Since this equation does not account for finite-range and intraspecies scattering effects, the result aligns reasonably well with our calculations.


	\section{Summary}
	In this paper, we studied the atom-dimer elastic scattering cross sections, scattering lengths and three-body recombination rates in $^{40}$K-$^{87}$Rb mixtures with total angular momentum $J=0$. We use the R-matrix propagation method in the hyperspherical coordinate frame based on the Lennard-Jones model potential. The finite intraspecies scattering length is considered with the $^{87}$Rb-$^{87}$Rb s-wave scattering length fixed at $a_{\scriptscriptstyle\textsl{RbRb}} = 100\,a_{0}$.
	
	To clarify whether there is an Efimov state disappearing at the $^{40}$K + $^{87}$Rb$^{87}$Rb threshold, we investigate the $^{40}$K + $^{87}$Rb$^{87}$Rb elastic scattering at a negative $^{87}$Rb-$^{40}$K scattering length. We calculate the scattering lengths and cross sections in the range from $-100\,a_{0} $ to $-1000\,a_{0}$. An atom-dimer elastic scattering resonance at $a_{\scriptscriptstyle\textsl{RbK}} = - 155\,a_0$ is found, which is due to the presence of the near-threshold trimer state at the atom-dimer threshold ($^{40}$K + $^{87}$Rb$^{87}$Rb). The three-body state is located at $|a_{\scriptscriptstyle\textsl{RbK}}|= 155\,a_{0}< 2 r_{\scriptscriptstyle \textsl{vdW,RbRb}} $, and its binding energy is larger than $E_{\scriptscriptstyle \textsl{vdW,RbRb}}$ by one order of magnitude. Thus, the three-body state is not an Efimov state, and it is not related to the previously observed atom-dimer Efimov resonance at $a_{\scriptscriptstyle*} = 230(30)\,a_{0}$. Additionally, a barrier exists in the lowest entrance channel on the negative $^{87}$Rb-$^{40}$K scattering length side, which could support a three-body shape resonance. The three-body resonance energies, $E_{R}$, and width, $\Gamma$, are obtained using the eigenphase sum method.
	
On the positive $a_{\scriptscriptstyle\textsl{RbK}}$ side, we identified an Efimov recombination minimum beyond the range of previous measurements. For $^{87}$Rb-$^{40}$K$^{87}$Rb inelastic scattering, a resonant feature was observed at $a_{\scriptscriptstyle *} = 133 \,a_{0}$ in the atom-dimer loss coefficients. Including the three-body potential brought the calculated atom-dimer resonance positions into agreement with experimental values. We labeled the present values of $a_{\scriptscriptstyle+}$, $a_{\scriptscriptstyle*}$, and $a_{\scriptscriptstyle *-}$ (the $^{40}$K-$^{87}$Rb$^{87}$Rb resonance position) in Figure\,\ref{1}. The combination of the Efimov feature with experimental observations of the atom-dimer Efimov resonance presents an opportunity to test the universality of Efimov features. Our study demonstrates the influence of finite-range effects and non-resonant intraspecies scattering length in $^{40}$K-$^{87}$Rb mixtures, helps elucidate the universality of three-body parameters for heteronuclear systems with finite intraspecies scattering lengths.

	\section{Acknowledgments}
	We thank C. H. Greene, Shuyu Zhou, Baochun Yang and Yu-Hsin Chen for helpful discussions. Hui-Li Han was supported by the National Natural Science Foundation of China under Grant Nos. 11874391. and 12274423. Ting-Yun Shi was supported by National Natural Science Foundation of China under Grants No. 12274423. All the calculations are done on the APM-Theoretical Computing Cluster(APM-TCC). Any data that support the findings of this study are included within the article.


\begin{thebibliography}{41}%
\makeatletter
\providecommand \@ifxundefined [1]{%
 \@ifx{#1\undefined}
}%
\providecommand \@ifnum [1]{%
 \ifnum #1\expandafter \@firstoftwo
 \else \expandafter \@secondoftwo
 \fi
}%
\providecommand \@ifx [1]{%
 \ifx #1\expandafter \@firstoftwo
 \else \expandafter \@secondoftwo
 \fi
}%
\providecommand \natexlab [1]{#1}%
\providecommand \enquote  [1]{``#1''}%
\providecommand \bibnamefont  [1]{#1}%
\providecommand \bibfnamefont [1]{#1}%
\providecommand \citenamefont [1]{#1}%
\providecommand \href@noop [0]{\@secondoftwo}%
\providecommand \href [0]{\begingroup \@sanitize@url \@href}%
\providecommand \@href[1]{\@@startlink{#1}\@@href}%
\providecommand \@@href[1]{\endgroup#1\@@endlink}%
\providecommand \@sanitize@url [0]{\catcode `\\12\catcode `\$12\catcode
  `\&12\catcode `\#12\catcode `\^12\catcode `\_12\catcode `\%12\relax}%
\providecommand \@@startlink[1]{}%
\providecommand \@@endlink[0]{}%
\providecommand \url  [0]{\begingroup\@sanitize@url \@url }%
\providecommand \@url [1]{\endgroup\@href {#1}{\urlprefix }}%
\providecommand \urlprefix  [0]{URL }%
\providecommand \Eprint [0]{\href }%
\providecommand \doibase [0]{http://dx.doi.org/}%
\providecommand \selectlanguage [0]{\@gobble}%
\providecommand \bibinfo  [0]{\@secondoftwo}%
\providecommand \bibfield  [0]{\@secondoftwo}%
\providecommand \translation [1]{[#1]}%
\providecommand \BibitemOpen [0]{}%
\providecommand \bibitemStop [0]{}%
\providecommand \bibitemNoStop [0]{.\EOS\space}%
\providecommand \EOS [0]{\spacefactor3000\relax}%
\providecommand \BibitemShut  [1]{\csname bibitem#1\endcsname}%
\let\auto@bib@innerbib\@empty
\bibitem [{\citenamefont {Efimov}(1970)}]{Efimov1970}%
  \BibitemOpen
  \bibfield  {author} {\bibinfo {author} {\bibfnamefont {V.}~\bibnamefont
  {Efimov}},\ }\href {\doibase https://doi.org/10.1016/0370-2693(70)90349-7}
  {\bibfield  {journal} {\bibinfo  {journal} {Phys. Lett. B}\ }\textbf
  {\bibinfo {volume} {33}},\ \bibinfo {pages} {563} (\bibinfo {year}
  {1970})}\BibitemShut {NoStop}%
\bibitem [{\citenamefont {Efimov}(1973)}]{Efimov1973}%
  \BibitemOpen
  \bibfield  {author} {\bibinfo {author} {\bibfnamefont {V.}~\bibnamefont
  {Efimov}},\ }\href {\doibase https://doi.org/10.1016/0375-9474(73)90510-1}
  {\bibfield  {journal} {\bibinfo  {journal} {Nucl. Phys. A}\ }\textbf
  {\bibinfo {volume} {210}},\ \bibinfo {pages} {157} (\bibinfo {year}
  {1973})}\BibitemShut {NoStop}%
\bibitem [{\citenamefont {Kraemer}\ \emph {et~al.}(2006)\citenamefont
  {Kraemer}, \citenamefont {Mark}, \citenamefont {Waldburger}, \citenamefont
  {Danzl}, \citenamefont {Chin}, \citenamefont {Engeser}, \citenamefont
  {Lange}, \citenamefont {Pilch}, \citenamefont {Jaakkola}, \citenamefont
  {N\"agerl},\ and\ \citenamefont {Grimm}}]{Kraemer2006}%
  \BibitemOpen
  \bibfield  {author} {\bibinfo {author} {\bibfnamefont {T.}~\bibnamefont
  {Kraemer}}, \bibinfo {author} {\bibfnamefont {M.}~\bibnamefont {Mark}},
  \bibinfo {author} {\bibfnamefont {P.}~\bibnamefont {Waldburger}}, \bibinfo
  {author} {\bibfnamefont {J.~G.}\ \bibnamefont {Danzl}}, \bibinfo {author}
  {\bibfnamefont {C.}~\bibnamefont {Chin}}, \bibinfo {author} {\bibfnamefont
  {B.}~\bibnamefont {Engeser}}, \bibinfo {author} {\bibfnamefont {A.~D.}\
  \bibnamefont {Lange}}, \bibinfo {author} {\bibfnamefont {K.}~\bibnamefont
  {Pilch}}, \bibinfo {author} {\bibfnamefont {A.}~\bibnamefont {Jaakkola}},
  \bibinfo {author} {\bibfnamefont {H.~C.}\ \bibnamefont {N\"agerl}}, \ and\
  \bibinfo {author} {\bibfnamefont {R.}~\bibnamefont {Grimm}},\ }\href
  {\doibase 10.1038/nature04626} {\bibfield  {journal} {\bibinfo  {journal}
  {Nature}\ }\textbf {\bibinfo {volume} {440}},\ \bibinfo {pages} {315–318}
  (\bibinfo {year} {2006})}\BibitemShut {NoStop}%
\bibitem [{\citenamefont {Zaccanti}\ \emph {et~al.}(2009)\citenamefont
  {Zaccanti}, \citenamefont {Deissler}, \citenamefont {D'Errico}, \citenamefont
  {Fattori}, \citenamefont {Jona-Lasinio}, \citenamefont {M\"uller},
  \citenamefont {Roati}, \citenamefont {Inguscio},\ and\ \citenamefont
  {Modugno}}]{Zaccanti2009}%
  \BibitemOpen
  \bibfield  {author} {\bibinfo {author} {\bibfnamefont {M.}~\bibnamefont
  {Zaccanti}}, \bibinfo {author} {\bibfnamefont {B.}~\bibnamefont {Deissler}},
  \bibinfo {author} {\bibfnamefont {C.}~\bibnamefont {D'Errico}}, \bibinfo
  {author} {\bibfnamefont {M.}~\bibnamefont {Fattori}}, \bibinfo {author}
  {\bibfnamefont {M.}~\bibnamefont {Jona-Lasinio}}, \bibinfo {author}
  {\bibfnamefont {S.}~\bibnamefont {M\"uller}}, \bibinfo {author}
  {\bibfnamefont {G.}~\bibnamefont {Roati}}, \bibinfo {author} {\bibfnamefont
  {M.}~\bibnamefont {Inguscio}}, \ and\ \bibinfo {author} {\bibfnamefont
  {G.}~\bibnamefont {Modugno}},\ }\href {\doibase 10.1038/nphys1334} {\bibfield
   {journal} {\bibinfo  {journal} {Nat. Phys.}\ }\textbf {\bibinfo {volume}
  {5}},\ \bibinfo {pages} {586} (\bibinfo {year} {2009})}\BibitemShut {NoStop}%
\bibitem [{\citenamefont {Bloom}\ \emph {et~al.}(2013)\citenamefont {Bloom},
  \citenamefont {Hu}, \citenamefont {Cumby},\ and\ \citenamefont
  {Jin}}]{Bloom2013}%
  \BibitemOpen
  \bibfield  {author} {\bibinfo {author} {\bibfnamefont {R.~S.}\ \bibnamefont
  {Bloom}}, \bibinfo {author} {\bibfnamefont {M.-G.}\ \bibnamefont {Hu}},
  \bibinfo {author} {\bibfnamefont {T.~D.}\ \bibnamefont {Cumby}}, \ and\
  \bibinfo {author} {\bibfnamefont {D.~S.}\ \bibnamefont {Jin}},\ }\href
  {\doibase 10.1103/PhysRevLett.111.105301} {\bibfield  {journal} {\bibinfo
  {journal} {Phys. Rev. Lett.}\ }\textbf {\bibinfo {volume} {111}},\ \bibinfo
  {pages} {105301} (\bibinfo {year} {2013})}\BibitemShut {NoStop}%
\bibitem [{\citenamefont {Ulmanis}\ \emph {et~al.}(2016)\citenamefont
  {Ulmanis}, \citenamefont {H\"afner}, \citenamefont {Pires}, \citenamefont
  {Kuhnle}, \citenamefont {Wang}, \citenamefont {Greene},\ and\ \citenamefont
  {Weidem\"uller}}]{HUlmanis2016}%
  \BibitemOpen
  \bibfield  {author} {\bibinfo {author} {\bibfnamefont {J.}~\bibnamefont
  {Ulmanis}}, \bibinfo {author} {\bibfnamefont {S.}~\bibnamefont {H\"afner}},
  \bibinfo {author} {\bibfnamefont {R.}~\bibnamefont {Pires}}, \bibinfo
  {author} {\bibfnamefont {E.~D.}\ \bibnamefont {Kuhnle}}, \bibinfo {author}
  {\bibfnamefont {Y.}~\bibnamefont {Wang}}, \bibinfo {author} {\bibfnamefont
  {C.~H.}\ \bibnamefont {Greene}}, \ and\ \bibinfo {author} {\bibfnamefont
  {M.}~\bibnamefont {Weidem\"uller}},\ }\href {\doibase
  10.1103/PhysRevLett.117.153201} {\bibfield  {journal} {\bibinfo  {journal}
  {Phys. Rev. Lett.}\ }\textbf {\bibinfo {volume} {117}},\ \bibinfo {pages}
  {153201} (\bibinfo {year} {2016})}\BibitemShut {NoStop}%
\bibitem [{\citenamefont {Barontini}\ \emph {et~al.}(2009)\citenamefont
  {Barontini}, \citenamefont {Weber}, \citenamefont {Rabatti}, \citenamefont
  {Catani}, \citenamefont {Thalhammer}, \citenamefont {Inguscio},\ and\
  \citenamefont {Minardi}}]{Barontini2009}%
  \BibitemOpen
  \bibfield  {author} {\bibinfo {author} {\bibfnamefont {G.}~\bibnamefont
  {Barontini}}, \bibinfo {author} {\bibfnamefont {C.}~\bibnamefont {Weber}},
  \bibinfo {author} {\bibfnamefont {F.}~\bibnamefont {Rabatti}}, \bibinfo
  {author} {\bibfnamefont {J.}~\bibnamefont {Catani}}, \bibinfo {author}
  {\bibfnamefont {G.}~\bibnamefont {Thalhammer}}, \bibinfo {author}
  {\bibfnamefont {M.}~\bibnamefont {Inguscio}}, \ and\ \bibinfo {author}
  {\bibfnamefont {F.}~\bibnamefont {Minardi}},\ }\href {\doibase
  10.1103/PhysRevLett.103.043201} {\bibfield  {journal} {\bibinfo  {journal}
  {Phys. Rev. Lett.}\ }\textbf {\bibinfo {volume} {103}},\ \bibinfo {pages}
  {043201} (\bibinfo {year} {2009})}\BibitemShut {NoStop}%
\bibitem [{\citenamefont {Jensen}\ \emph {et~al.}(2004)\citenamefont {Jensen},
  \citenamefont {Riisager}, \citenamefont {Fedorov},\ and\ \citenamefont
  {Garrido}}]{Jensen2004structure}%
  \BibitemOpen
  \bibfield  {author} {\bibinfo {author} {\bibfnamefont {A.~S.}\ \bibnamefont
  {Jensen}}, \bibinfo {author} {\bibfnamefont {K.}~\bibnamefont {Riisager}},
  \bibinfo {author} {\bibfnamefont {D.~V.}\ \bibnamefont {Fedorov}}, \ and\
  \bibinfo {author} {\bibfnamefont {E.}~\bibnamefont {Garrido}},\ }\href
  {\doibase 10.1103/RevModPhys.76.215} {\bibfield  {journal} {\bibinfo
  {journal} {Rev. Mod. Phys.}\ }\textbf {\bibinfo {volume} {76}},\ \bibinfo
  {pages} {215} (\bibinfo {year} {2004})}\BibitemShut {NoStop}%
\bibitem [{\citenamefont {Kolganova}\ and\ \citenamefont
  {Sandhas}(2011)}]{Kolganova2011}%
  \BibitemOpen
  \bibfield  {author} {\bibinfo {author} {\bibfnamefont {M.~A.~K.}\
  \bibnamefont {Kolganova}, \bibfnamefont {E.~A.}}\ and\ \bibinfo {author}
  {\bibfnamefont {W.}~\bibnamefont {Sandhas}},\ }\href {\doibase
  10.1007/s00601-011-0233-x} {\bibfield  {journal} {\bibinfo  {journal}
  {Few-Body Syst.}\ }\textbf {\bibinfo {volume} {51}},\ \bibinfo {pages} {249}
  (\bibinfo {year} {2011})}\BibitemShut {NoStop}%
\bibitem [{\citenamefont {Naidon}\ and\ \citenamefont
  {Endo}(2017)}]{Naidon2017}%
  \BibitemOpen
  \bibfield  {author} {\bibinfo {author} {\bibfnamefont {P.}~\bibnamefont
  {Naidon}}\ and\ \bibinfo {author} {\bibfnamefont {S.}~\bibnamefont {Endo}},\
  }\href {\doibase 10.1088/1361-6633/aa50e8} {\bibfield  {journal} {\bibinfo
  {journal} {Rep. Prog. Phys.}\ }\textbf {\bibinfo {volume} {80}},\ \bibinfo
  {pages} {056001} (\bibinfo {year} {2017})}\BibitemShut {NoStop}%
\bibitem [{\citenamefont {Braaten}\ and\ \citenamefont
  {Hammer}(2006)}]{Eric2006}%
  \BibitemOpen
  \bibfield  {author} {\bibinfo {author} {\bibfnamefont {E.}~\bibnamefont
  {Braaten}}\ and\ \bibinfo {author} {\bibfnamefont {H.-W.}\ \bibnamefont
  {Hammer}},\ }\href {\doibase https://doi.org/10.1016/j.physrep.2006.03.001}
  {\bibfield  {journal} {\bibinfo  {journal} {Phys. Rep.}\ }\textbf {\bibinfo
  {volume} {428}},\ \bibinfo {pages} {259} (\bibinfo {year}
  {2006})}\BibitemShut {NoStop}%
\bibitem [{\citenamefont {D'Incao}\ and\ \citenamefont
  {Esry}(2006)}]{IncaoMass2006}%
  \BibitemOpen
  \bibfield  {author} {\bibinfo {author} {\bibfnamefont {J.~P.}\ \bibnamefont
  {D'Incao}}\ and\ \bibinfo {author} {\bibfnamefont {B.~D.}\ \bibnamefont
  {Esry}},\ }\href {\doibase 10.1103/PhysRevA.73.030702} {\bibfield  {journal}
  {\bibinfo  {journal} {Phys. Rev. A}\ }\textbf {\bibinfo {volume} {73}},\
  \bibinfo {pages} {030702} (\bibinfo {year} {2006})}\BibitemShut {NoStop}%
\bibitem [{\citenamefont {Chin}\ \emph {et~al.}(2010)\citenamefont {Chin},
  \citenamefont {Grimm}, \citenamefont {Julienne},\ and\ \citenamefont
  {Tiesinga}}]{Chin2010}%
  \BibitemOpen
  \bibfield  {author} {\bibinfo {author} {\bibfnamefont {C.}~\bibnamefont
  {Chin}}, \bibinfo {author} {\bibfnamefont {R.}~\bibnamefont {Grimm}},
  \bibinfo {author} {\bibfnamefont {P.}~\bibnamefont {Julienne}}, \ and\
  \bibinfo {author} {\bibfnamefont {E.}~\bibnamefont {Tiesinga}},\ }\href
  {\doibase 10.1103/RevModPhys.82.1225} {\bibfield  {journal} {\bibinfo
  {journal} {Rev. Mod. Phys.}\ }\textbf {\bibinfo {volume} {82}},\ \bibinfo
  {pages} {1225} (\bibinfo {year} {2010})}\BibitemShut {NoStop}%
\bibitem [{\citenamefont {D'Incao}(2018)}]{D_Incao_2018}%
  \BibitemOpen
  \bibfield  {author} {\bibinfo {author} {\bibfnamefont {J.~P.}\ \bibnamefont
  {D'Incao}},\ }\href {\doibase 10.1088/1361-6455/aaa116} {\bibfield  {journal}
  {\bibinfo  {journal} {J. Phys. B: At. Mol. Opt. Phys.}\ }\textbf {\bibinfo
  {volume} {51}},\ \bibinfo {pages} {043001} (\bibinfo {year}
  {2018})}\BibitemShut {NoStop}%
\bibitem [{\citenamefont {Maier}\ \emph {et~al.}(2015)\citenamefont {Maier},
  \citenamefont {Eisele}, \citenamefont {Tiemann},\ and\ \citenamefont
  {Zimmermann}}]{Maier2015}%
  \BibitemOpen
  \bibfield  {author} {\bibinfo {author} {\bibfnamefont {R.~A.~W.}\
  \bibnamefont {Maier}}, \bibinfo {author} {\bibfnamefont {M.}~\bibnamefont
  {Eisele}}, \bibinfo {author} {\bibfnamefont {E.}~\bibnamefont {Tiemann}}, \
  and\ \bibinfo {author} {\bibfnamefont {C.}~\bibnamefont {Zimmermann}},\
  }\href {\doibase 10.1103/PhysRevLett.115.043201} {\bibfield  {journal}
  {\bibinfo  {journal} {Phys. Rev. Lett.}\ }\textbf {\bibinfo {volume} {115}},\
  \bibinfo {pages} {043201} (\bibinfo {year} {2015})}\BibitemShut {NoStop}%
\bibitem [{\citenamefont {Tung}\ \emph {et~al.}(2014)\citenamefont {Tung},
  \citenamefont {Jim\'enez-Garc\'{\i}a}, \citenamefont {Johansen},
  \citenamefont {Parker},\ and\ \citenamefont {Chin}}]{Tung2014Dec}%
  \BibitemOpen
  \bibfield  {author} {\bibinfo {author} {\bibfnamefont {S.-K.}\ \bibnamefont
  {Tung}}, \bibinfo {author} {\bibfnamefont {K.}~\bibnamefont
  {Jim\'enez-Garc\'{\i}a}}, \bibinfo {author} {\bibfnamefont {J.}~\bibnamefont
  {Johansen}}, \bibinfo {author} {\bibfnamefont {C.~V.}\ \bibnamefont
  {Parker}}, \ and\ \bibinfo {author} {\bibfnamefont {C.}~\bibnamefont
  {Chin}},\ }\href {\doibase 10.1103/PhysRevLett.113.240402} {\bibfield
  {journal} {\bibinfo  {journal} {Phys. Rev. Lett.}\ }\textbf {\bibinfo
  {volume} {113}},\ \bibinfo {pages} {240402} (\bibinfo {year}
  {2014})}\BibitemShut {NoStop}%
\bibitem [{\citenamefont {Huang}\ \emph {et~al.}(2014)\citenamefont {Huang},
  \citenamefont {Sidorenkov}, \citenamefont {Grimm},\ and\ \citenamefont
  {Hutson}}]{Huang2014}%
  \BibitemOpen
  \bibfield  {author} {\bibinfo {author} {\bibfnamefont {B.}~\bibnamefont
  {Huang}}, \bibinfo {author} {\bibfnamefont {L.~A.}\ \bibnamefont
  {Sidorenkov}}, \bibinfo {author} {\bibfnamefont {R.}~\bibnamefont {Grimm}}, \
  and\ \bibinfo {author} {\bibfnamefont {J.~M.}\ \bibnamefont {Hutson}},\
  }\href {\doibase 10.1103/PhysRevLett.112.190401} {\bibfield  {journal}
  {\bibinfo  {journal} {Phys. Rev. Lett.}\ }\textbf {\bibinfo {volume} {112}},\
  \bibinfo {pages} {190401} (\bibinfo {year} {2014})}\BibitemShut {NoStop}%
\bibitem [{\citenamefont {H\"afner}\ \emph {et~al.}(2017)\citenamefont
  {H\"afner}, \citenamefont {Ulmanis}, \citenamefont {Kuhnle}, \citenamefont
  {Wang}, \citenamefont {Greene},\ and\ \citenamefont
  {Weidem\"uller}}]{hafner2017role}%
  \BibitemOpen
  \bibfield  {author} {\bibinfo {author} {\bibfnamefont {S.}~\bibnamefont
  {H\"afner}}, \bibinfo {author} {\bibfnamefont {J.}~\bibnamefont {Ulmanis}},
  \bibinfo {author} {\bibfnamefont {E.~D.}\ \bibnamefont {Kuhnle}}, \bibinfo
  {author} {\bibfnamefont {Y.}~\bibnamefont {Wang}}, \bibinfo {author}
  {\bibfnamefont {C.~H.}\ \bibnamefont {Greene}}, \ and\ \bibinfo {author}
  {\bibfnamefont {M.}~\bibnamefont {Weidem\"uller}},\ }\href {\doibase
  10.1103/PhysRevA.95.062708} {\bibfield  {journal} {\bibinfo  {journal} {Phys.
  Rev. A}\ }\textbf {\bibinfo {volume} {95}},\ \bibinfo {pages} {062708}
  (\bibinfo {year} {2017})}\BibitemShut {NoStop}%
\bibitem [{\citenamefont {D’Incao}\ \emph {et~al.}(2004)\citenamefont
  {D’Incao}, \citenamefont {Suno},\ and\ \citenamefont {Esry}}]{d2004limits}%
  \BibitemOpen
  \bibfield  {author} {\bibinfo {author} {\bibfnamefont {J.~P.}\ \bibnamefont
  {D’Incao}}, \bibinfo {author} {\bibfnamefont {H.}~\bibnamefont {Suno}}, \
  and\ \bibinfo {author} {\bibfnamefont {B.~D.}\ \bibnamefont {Esry}},\ }\href
  {\doibase https://doi.org/10.1103/PhysRevLett.93.123201} {\bibfield
  {journal} {\bibinfo  {journal} {Phys. Rev. Lett.}\ }\textbf {\bibinfo
  {volume} {93}},\ \bibinfo {pages} {123201} (\bibinfo {year}
  {2004})}\BibitemShut {NoStop}%
\bibitem [{\citenamefont {Wang}\ \emph
  {et~al.}(2012{\natexlab{a}})\citenamefont {Wang}, \citenamefont {D'Incao},
  \citenamefont {Esry},\ and\ \citenamefont {Greene}}]{WangJia2012}%
  \BibitemOpen
  \bibfield  {author} {\bibinfo {author} {\bibfnamefont {J.}~\bibnamefont
  {Wang}}, \bibinfo {author} {\bibfnamefont {J.~P.}\ \bibnamefont {D'Incao}},
  \bibinfo {author} {\bibfnamefont {B.~D.}\ \bibnamefont {Esry}}, \ and\
  \bibinfo {author} {\bibfnamefont {C.~H.}\ \bibnamefont {Greene}},\ }\href
  {\doibase 10.1103/PhysRevLett.108.263001} {\bibfield  {journal} {\bibinfo
  {journal} {Phys. Rev. Lett.}\ }\textbf {\bibinfo {volume} {108}},\ \bibinfo
  {pages} {263001} (\bibinfo {year} {2012}{\natexlab{a}})}\BibitemShut
  {NoStop}%
\bibitem [{\citenamefont {Naidon}\ \emph
  {et~al.}(2014{\natexlab{a}})\citenamefont {Naidon}, \citenamefont {Endo},\
  and\ \citenamefont {Ueda}}]{NaidonAug2014}%
  \BibitemOpen
  \bibfield  {author} {\bibinfo {author} {\bibfnamefont {P.}~\bibnamefont
  {Naidon}}, \bibinfo {author} {\bibfnamefont {S.}~\bibnamefont {Endo}}, \ and\
  \bibinfo {author} {\bibfnamefont {M.}~\bibnamefont {Ueda}},\ }\href {\doibase
  10.1103/PhysRevA.90.022106} {\bibfield  {journal} {\bibinfo  {journal} {Phys.
  Rev. A}\ }\textbf {\bibinfo {volume} {90}},\ \bibinfo {pages} {022106}
  (\bibinfo {year} {2014}{\natexlab{a}})}\BibitemShut {NoStop}%
\bibitem [{\citenamefont {Naidon}\ \emph
  {et~al.}(2014{\natexlab{b}})\citenamefont {Naidon}, \citenamefont {Endo},\
  and\ \citenamefont {Ueda}}]{NaidonMar2014}%
  \BibitemOpen
  \bibfield  {author} {\bibinfo {author} {\bibfnamefont {P.}~\bibnamefont
  {Naidon}}, \bibinfo {author} {\bibfnamefont {S.}~\bibnamefont {Endo}}, \ and\
  \bibinfo {author} {\bibfnamefont {M.}~\bibnamefont {Ueda}},\ }\href {\doibase
  10.1103/PhysRevLett.112.105301} {\bibfield  {journal} {\bibinfo  {journal}
  {Phys. Rev. Lett.}\ }\textbf {\bibinfo {volume} {112}},\ \bibinfo {pages}
  {105301} (\bibinfo {year} {2014}{\natexlab{b}})}\BibitemShut {NoStop}%
\bibitem [{\citenamefont {Gogolin}\ \emph {et~al.}(2008)\citenamefont
  {Gogolin}, \citenamefont {Mora},\ and\ \citenamefont
  {Egger}}]{Gogolin2008Apr}%
  \BibitemOpen
  \bibfield  {author} {\bibinfo {author} {\bibfnamefont {A.~O.}\ \bibnamefont
  {Gogolin}}, \bibinfo {author} {\bibfnamefont {C.}~\bibnamefont {Mora}}, \
  and\ \bibinfo {author} {\bibfnamefont {R.}~\bibnamefont {Egger}},\ }\href
  {\doibase 10.1103/PhysRevLett.100.140404} {\bibfield  {journal} {\bibinfo
  {journal} {Phys. Rev. Lett.}\ }\textbf {\bibinfo {volume} {100}},\ \bibinfo
  {pages} {140404} (\bibinfo {year} {2008})}\BibitemShut {NoStop}%
\bibitem [{\citenamefont {Mestrom}\ \emph {et~al.}(2017)\citenamefont
  {Mestrom}, \citenamefont {Wang}, \citenamefont {Greene},\ and\ \citenamefont
  {D'Incao}}]{mestrom2017efimov}%
  \BibitemOpen
  \bibfield  {author} {\bibinfo {author} {\bibfnamefont {P.~M.~A.}\
  \bibnamefont {Mestrom}}, \bibinfo {author} {\bibfnamefont {J.}~\bibnamefont
  {Wang}}, \bibinfo {author} {\bibfnamefont {C.~H.}\ \bibnamefont {Greene}}, \
  and\ \bibinfo {author} {\bibfnamefont {J.~P.}\ \bibnamefont {D'Incao}},\
  }\href {\doibase 10.1103/PhysRevA.95.032707} {\bibfield  {journal} {\bibinfo
  {journal} {Phys. Rev. A}\ }\textbf {\bibinfo {volume} {95}},\ \bibinfo
  {pages} {032707} (\bibinfo {year} {2017})}\BibitemShut {NoStop}%
\bibitem [{\citenamefont {Wacker}\ \emph {et~al.}(2016)\citenamefont {Wacker},
  \citenamefont {J\o{}rgensen}, \citenamefont {Birkmose}, \citenamefont
  {Winter}, \citenamefont {Mikkelsen}, \citenamefont {Sherson}, \citenamefont
  {Zinner},\ and\ \citenamefont {Arlt}}]{Wacker2016KRb}%
  \BibitemOpen
  \bibfield  {author} {\bibinfo {author} {\bibfnamefont {L.~J.}\ \bibnamefont
  {Wacker}}, \bibinfo {author} {\bibfnamefont {N.~B.}\ \bibnamefont
  {J\o{}rgensen}}, \bibinfo {author} {\bibfnamefont {D.}~\bibnamefont
  {Birkmose}}, \bibinfo {author} {\bibfnamefont {N.}~\bibnamefont {Winter}},
  \bibinfo {author} {\bibfnamefont {M.}~\bibnamefont {Mikkelsen}}, \bibinfo
  {author} {\bibfnamefont {J.}~\bibnamefont {Sherson}}, \bibinfo {author}
  {\bibfnamefont {N.}~\bibnamefont {Zinner}}, \ and\ \bibinfo {author}
  {\bibfnamefont {J.~J.}\ \bibnamefont {Arlt}},\ }\href {\doibase
  10.1103/PhysRevLett.117.163201} {\bibfield  {journal} {\bibinfo  {journal}
  {Phys. Rev. Lett.}\ }\textbf {\bibinfo {volume} {117}},\ \bibinfo {pages}
  {163201} (\bibinfo {year} {2016})}\BibitemShut {NoStop}%
\bibitem [{\citenamefont {Wang}\ \emph
  {et~al.}(2012{\natexlab{b}})\citenamefont {Wang}, \citenamefont {Wang},
  \citenamefont {D'Incao},\ and\ \citenamefont {Greene}}]{WangYujun2012}%
  \BibitemOpen
  \bibfield  {author} {\bibinfo {author} {\bibfnamefont {Y.}~\bibnamefont
  {Wang}}, \bibinfo {author} {\bibfnamefont {J.}~\bibnamefont {Wang}}, \bibinfo
  {author} {\bibfnamefont {J.~P.}\ \bibnamefont {D'Incao}}, \ and\ \bibinfo
  {author} {\bibfnamefont {C.~H.}\ \bibnamefont {Greene}},\ }\href {\doibase
  10.1103/PhysRevLett.109.243201} {\bibfield  {journal} {\bibinfo  {journal}
  {Phys. Rev. Lett.}\ }\textbf {\bibinfo {volume} {109}},\ \bibinfo {pages}
  {243201} (\bibinfo {year} {2012}{\natexlab{b}})}\BibitemShut {NoStop}%
\bibitem [{\citenamefont {Helfrich}\ \emph {et~al.}(2010)\citenamefont
  {Helfrich}, \citenamefont {Hammer},\ and\ \citenamefont
  {Petrov}}]{Helfrich2010}%
  \BibitemOpen
  \bibfield  {author} {\bibinfo {author} {\bibfnamefont {K.}~\bibnamefont
  {Helfrich}}, \bibinfo {author} {\bibfnamefont {H.-W.}\ \bibnamefont
  {Hammer}}, \ and\ \bibinfo {author} {\bibfnamefont {D.~S.}\ \bibnamefont
  {Petrov}},\ }\href {\doibase 10.1103/PhysRevA.81.042715} {\bibfield
  {journal} {\bibinfo  {journal} {Phys. Rev. A}\ }\textbf {\bibinfo {volume}
  {81}},\ \bibinfo {pages} {042715} (\bibinfo {year} {2010})}\BibitemShut
  {NoStop}%
\bibitem [{\citenamefont {Hu}\ \emph {et~al.}(2014)\citenamefont {Hu},
  \citenamefont {Bloom}, \citenamefont {Jin},\ and\ \citenamefont
  {Goldwin}}]{Hu2014}%
  \BibitemOpen
  \bibfield  {author} {\bibinfo {author} {\bibfnamefont {M.-G.}\ \bibnamefont
  {Hu}}, \bibinfo {author} {\bibfnamefont {R.~S.}\ \bibnamefont {Bloom}},
  \bibinfo {author} {\bibfnamefont {D.~S.}\ \bibnamefont {Jin}}, \ and\
  \bibinfo {author} {\bibfnamefont {J.~M.}\ \bibnamefont {Goldwin}},\ }\href
  {\doibase 10.1103/PhysRevA.90.013619} {\bibfield  {journal} {\bibinfo
  {journal} {Phys. Rev. A}\ }\textbf {\bibinfo {volume} {90}},\ \bibinfo
  {pages} {013619} (\bibinfo {year} {2014})}\BibitemShut {NoStop}%
\bibitem [{\citenamefont {Zirbel}\ \emph {et~al.}(2008)\citenamefont {Zirbel},
  \citenamefont {Ni}, \citenamefont {Ospelkaus}, \citenamefont {D'Incao},
  \citenamefont {Wieman}, \citenamefont {Ye},\ and\ \citenamefont
  {Jin}}]{Zirbel2008}%
  \BibitemOpen
  \bibfield  {author} {\bibinfo {author} {\bibfnamefont {J.~J.}\ \bibnamefont
  {Zirbel}}, \bibinfo {author} {\bibfnamefont {K.-K.}\ \bibnamefont {Ni}},
  \bibinfo {author} {\bibfnamefont {S.}~\bibnamefont {Ospelkaus}}, \bibinfo
  {author} {\bibfnamefont {J.~P.}\ \bibnamefont {D'Incao}}, \bibinfo {author}
  {\bibfnamefont {C.~E.}\ \bibnamefont {Wieman}}, \bibinfo {author}
  {\bibfnamefont {J.}~\bibnamefont {Ye}}, \ and\ \bibinfo {author}
  {\bibfnamefont {D.~S.}\ \bibnamefont {Jin}},\ }\href {\doibase
  10.1103/PhysRevLett.100.143201} {\bibfield  {journal} {\bibinfo  {journal}
  {Phys. Rev. Lett.}\ }\textbf {\bibinfo {volume} {100}},\ \bibinfo {pages}
  {143201} (\bibinfo {year} {2008})}\BibitemShut {NoStop}%
\bibitem [{\citenamefont {Wang}\ \emph {et~al.}(2011)\citenamefont {Wang},
  \citenamefont {D'Incao},\ and\ \citenamefont {Greene}}]{WangJia2011}%
  \BibitemOpen
  \bibfield  {author} {\bibinfo {author} {\bibfnamefont {J.}~\bibnamefont
  {Wang}}, \bibinfo {author} {\bibfnamefont {J.~P.}\ \bibnamefont {D'Incao}}, \
  and\ \bibinfo {author} {\bibfnamefont {C.~H.}\ \bibnamefont {Greene}},\
  }\href {\doibase 10.1103/PhysRevA.84.052721} {\bibfield  {journal} {\bibinfo
  {journal} {Phys. Rev. A}\ }\textbf {\bibinfo {volume} {84}},\ \bibinfo
  {pages} {052721} (\bibinfo {year} {2011})}\BibitemShut {NoStop}%
\bibitem [{\citenamefont {Tolstikhin}\ \emph {et~al.}(1996)\citenamefont
  {Tolstikhin}, \citenamefont {Watanabe},\ and\ \citenamefont
  {Matsuzawa}}]{Tolstikhin1996SVD}%
  \BibitemOpen
  \bibfield  {author} {\bibinfo {author} {\bibfnamefont {O.~I.}\ \bibnamefont
  {Tolstikhin}}, \bibinfo {author} {\bibfnamefont {S.}~\bibnamefont
  {Watanabe}}, \ and\ \bibinfo {author} {\bibfnamefont {M.}~\bibnamefont
  {Matsuzawa}},\ }\href {\doibase 10.1088/0953-4075/29/11/001} {\bibfield
  {journal} {\bibinfo  {journal} {J. Phys. B: At. Mol. Opt. Phys.}\ }\textbf
  {\bibinfo {volume} {29}},\ \bibinfo {pages} {L389} (\bibinfo {year}
  {1996})}\BibitemShut {NoStop}%
\bibitem [{\citenamefont {Lin}(1995)}]{lin1995}%
  \BibitemOpen
  \bibfield  {author} {\bibinfo {author} {\bibfnamefont {C.~D.}\ \bibnamefont
  {Lin}},\ }\href {\doibase https://doi.org/10.1016/0370-1573(94)00094-J}
  {\bibfield  {journal} {\bibinfo  {journal} {Phys. Rep.}\ }\textbf {\bibinfo
  {volume} {257}},\ \bibinfo {pages} {1} (\bibinfo {year} {1995})}\BibitemShut
  {NoStop}%
\bibitem [{\citenamefont {Porsev}\ \emph {et~al.}(2014)\citenamefont {Porsev},
  \citenamefont {Safronova}, \citenamefont {Derevianko},\ and\ \citenamefont
  {Clark}}]{C6RbRb2014}%
  \BibitemOpen
  \bibfield  {author} {\bibinfo {author} {\bibfnamefont {S.~G.}\ \bibnamefont
  {Porsev}}, \bibinfo {author} {\bibfnamefont {M.~S.}\ \bibnamefont
  {Safronova}}, \bibinfo {author} {\bibfnamefont {A.}~\bibnamefont
  {Derevianko}}, \ and\ \bibinfo {author} {\bibfnamefont {C.~W.}\ \bibnamefont
  {Clark}},\ }\href {\doibase 10.1103/PhysRevA.89.022703} {\bibfield  {journal}
  {\bibinfo  {journal} {Phys. Rev. A}\ }\textbf {\bibinfo {volume} {89}},\
  \bibinfo {pages} {022703} (\bibinfo {year} {2014})}\BibitemShut {NoStop}%
\bibitem [{\citenamefont {Marinescu}\ and\ \citenamefont
  {Sadeghpour}(1999)}]{C6RbK1999}%
  \BibitemOpen
  \bibfield  {author} {\bibinfo {author} {\bibfnamefont {M.}~\bibnamefont
  {Marinescu}}\ and\ \bibinfo {author} {\bibfnamefont {H.~R.}\ \bibnamefont
  {Sadeghpour}},\ }\href {\doibase 10.1103/PhysRevA.59.390} {\bibfield
  {journal} {\bibinfo  {journal} {Phys. Rev. A}\ }\textbf {\bibinfo {volume}
  {59}},\ \bibinfo {pages} {390} (\bibinfo {year} {1999})}\BibitemShut
  {NoStop}%
\bibitem [{\citenamefont {Zhao}\ \emph {et~al.}(2022)\citenamefont {Zhao},
  \citenamefont {Zhang}, \citenamefont {Han},\ and\ \citenamefont
  {Shi}}]{cyzhao2022}%
  \BibitemOpen
  \bibfield  {author} {\bibinfo {author} {\bibfnamefont {C.-Y.}\ \bibnamefont
  {Zhao}}, \bibinfo {author} {\bibfnamefont {Y.}~\bibnamefont {Zhang}},
  \bibinfo {author} {\bibfnamefont {H.-L.}\ \bibnamefont {Han}}, \ and\
  \bibinfo {author} {\bibfnamefont {T.-Y.}\ \bibnamefont {Shi}},\ }\href
  {\doibase 10.1007/s00601-022-01775-9} {\bibfield  {journal} {\bibinfo
  {journal} {Few-Body Syst.}\ }\textbf {\bibinfo {volume} {63}},\ \bibinfo
  {pages} {75} (\bibinfo {year} {2022})}\BibitemShut {NoStop}%
\bibitem [{\citenamefont {Esry}\ \emph {et~al.}(2001)\citenamefont {Esry},
  \citenamefont {Greene},\ and\ \citenamefont {Suno}}]{Esry2001Dec}%
  \BibitemOpen
  \bibfield  {author} {\bibinfo {author} {\bibfnamefont {B.~D.}\ \bibnamefont
  {Esry}}, \bibinfo {author} {\bibfnamefont {C.~H.}\ \bibnamefont {Greene}}, \
  and\ \bibinfo {author} {\bibfnamefont {H.}~\bibnamefont {Suno}},\ }\href
  {\doibase 10.1103/PhysRevA.65.010705} {\bibfield  {journal} {\bibinfo
  {journal} {Phys. Rev. A}\ }\textbf {\bibinfo {volume} {65}},\ \bibinfo
  {pages} {010705} (\bibinfo {year} {2001})}\BibitemShut {NoStop}%
\bibitem [{\citenamefont {D'Incao}\ and\ \citenamefont
  {Esry}(2005)}]{Incao2005Jun}%
  \BibitemOpen
  \bibfield  {author} {\bibinfo {author} {\bibfnamefont {J.~P.}\ \bibnamefont
  {D'Incao}}\ and\ \bibinfo {author} {\bibfnamefont {B.~D.}\ \bibnamefont
  {Esry}},\ }\href {\doibase 10.1103/PhysRevLett.94.213201} {\bibfield
  {journal} {\bibinfo  {journal} {Phys. Rev. Lett.}\ }\textbf {\bibinfo
  {volume} {94}},\ \bibinfo {pages} {213201} (\bibinfo {year}
  {2005})}\BibitemShut {NoStop}%
\bibitem [{\citenamefont {Gong}\ \emph {et~al.}(2019)\citenamefont {Gong},
  \citenamefont {Sun}, \citenamefont {Wang},\ and\ \citenamefont
  {Zhou}}]{Gong2019}%
  \BibitemOpen
  \bibfield  {author} {\bibinfo {author} {\bibfnamefont {D.}~\bibnamefont
  {Gong}}, \bibinfo {author} {\bibfnamefont {Z.}~\bibnamefont {Sun}}, \bibinfo
  {author} {\bibfnamefont {Y.}~\bibnamefont {Wang}}, \ and\ \bibinfo {author}
  {\bibfnamefont {S.}~\bibnamefont {Zhou}},\ }\href {\doibase
  10.1063/1.5130931} {\bibfield  {journal} {\bibinfo  {journal} {AIP Advances}\
  }\textbf {\bibinfo {volume} {9}},\ \bibinfo {pages} {125138} (\bibinfo {year}
  {2019})}\BibitemShut {NoStop}%
\bibitem [{\citenamefont {Pollack}\ \emph {et~al.}(2009)\citenamefont
  {Pollack}, \citenamefont {Dries},\ and\ \citenamefont {Hulet}}]{Pollack2009}%
  \BibitemOpen
  \bibfield  {author} {\bibinfo {author} {\bibfnamefont {S.~E.}\ \bibnamefont
  {Pollack}}, \bibinfo {author} {\bibfnamefont {D.}~\bibnamefont {Dries}}, \
  and\ \bibinfo {author} {\bibfnamefont {R.~G.}\ \bibnamefont {Hulet}},\ }\href
  {\doibase 10.1126/science.1182840} {\bibfield  {journal} {\bibinfo  {journal}
  {Science}\ }\textbf {\bibinfo {volume} {326}},\ \bibinfo {pages} {1683}
  (\bibinfo {year} {2009})}\BibitemShut {NoStop}%
\bibitem [{\citenamefont {Yudkin}\ \emph {et~al.}(2024)\citenamefont {Yudkin},
  \citenamefont {Elbaz}, \citenamefont {D’Incao}, \citenamefont {Julienne},\
  and\ \citenamefont {Khaykovich}}]{Yudkin2024}%
  \BibitemOpen
  \bibfield  {author} {\bibinfo {author} {\bibfnamefont {Y.}~\bibnamefont
  {Yudkin}}, \bibinfo {author} {\bibfnamefont {R.}~\bibnamefont {Elbaz}},
  \bibinfo {author} {\bibfnamefont {J.}~\bibnamefont {D’Incao}}, \bibinfo
  {author} {\bibfnamefont {P.~S.}\ \bibnamefont {Julienne}}, \ and\ \bibinfo
  {author} {\bibfnamefont {L.}~\bibnamefont {Khaykovich}},\ }\href {\doibase
  10.1038/s41467-024-46353-1} {\bibfield  {journal} {\bibinfo  {journal} {Nat.
  Commun.}\ }\textbf {\bibinfo {volume} {15}},\ \bibinfo {pages} {2127}
  (\bibinfo {year} {2024})}\BibitemShut {NoStop}%
\bibitem [{\citenamefont {Acharya}\ \emph {et~al.}(2016)\citenamefont
  {Acharya}, \citenamefont {Ji},\ and\ \citenamefont {Platter}}]{Acharya2016}%
  \BibitemOpen
  \bibfield  {author} {\bibinfo {author} {\bibfnamefont {B.}~\bibnamefont
  {Acharya}}, \bibinfo {author} {\bibfnamefont {C.}~\bibnamefont {Ji}}, \ and\
  \bibinfo {author} {\bibfnamefont {L.}~\bibnamefont {Platter}},\ }\href
  {\doibase 10.1103/PhysRevA.94.032702} {\bibfield  {journal} {\bibinfo
  {journal} {Phys. Rev. A}\ }\textbf {\bibinfo {volume} {94}},\ \bibinfo
  {pages} {032702} (\bibinfo {year} {2016})}\BibitemShut {NoStop}%
\end{thebibliography}%
	%

\end{document}